\documentclass[twocolumn]{aastex631}
\usepackage[utf8]{inputenc}  
\usepackage{textcomp}        
\usepackage[english]{babel}
\usepackage{url}
\usepackage{wrapfig}

\usepackage{fancyhdr}
\usepackage{color}
\usepackage{enumitem}
\usepackage{subfigure}
\usepackage{graphicx}

\usepackage[dvipsnames]{xcolor}
\usepackage{tcolorbox}
\usepackage{ragged2e}
\usepackage{xspace}
\usepackage{booktabs}
\let\tablenum\relax
\usepackage{siunitx}
\usepackage{tabularx}
\usepackage{scrextend}

\newcommand{\um}{\text{\,\textmu m\xspace}}
\newcommand{\teff}{$\text{T}_\mathrm{eff}$}
\newcommand{\logg}{log\,($\text{g}$)}

\newcommand{\mjup}{$\text{M}_\mathrm{J}$}

\newcommand{\rjup}{$\text{R}_\mathrm{J}$}
\newcommand{\rsun}{$\text{R}_\mathrm{\odot}$}
\newcommand{\lsun}{$\text{L}_\mathrm{\odot}$}

\newcommand{\jhu}{Department of Physics $\&$ Astronomy, Johns Hopkins University, 3400 N. Charles Street, Baltimore, MD 21218, USA}
\newcommand{\stsci}{Space Telescope Science Institute, 3700 San Martin Dr, Baltimore, MD 21218, USA}
\newcommand{\uofm}{Department of Astronomy, University of Michigan, 1085 S. University Ave, Ann Arbor, MI 48109, USA}
\newcommand{\IAedinburgh}{Institute for Astronomy, University of Edinburgh, Royal Observatory, Blackford Hill, Edinburgh, EH9 3HJ, UK}
\newcommand{\CESedinburgh}{Centre for Exoplanet Science, University of Edinburgh, Edinburgh, EH9 3HJ, UK}
\newcommand{\exeter}{Department of Physics and Astronomy, University of Exeter, Stocker Road, Exeter EX4 4QL, UK}
\newcommand{\marseille}{Laboratoire d'Astrophysique de Marseille, 38 Rue Fréderic Joliot Curie, 130013 Marseille, France}
\newcommand{\toronto}{Department of Astronomy and Astrophysics, University of Toronto, 50 St. George Street, Toronto, ON M5S 3H4, Canada}
\newcommand{\mitkavli}{MIT Kavli Institute for Astrophysics and Space Research, Massachusetts Institute of Technology, Cambridge, MA 02139, USA}
\newcommand{\UnivDiegoPortales}{Instituto de Estudios Astrof\'isicos, Facultad de Ingenier\'ia y Ciencias, Universidad Diego Portales, Av. Ej\'ercito Libertador 441, Santiago, Chile}
\newcommand{\yems}{Millennium Nucleus on Young Exoplanets and their Moons (YEMS)} 
\newcommand{\padova}{INAF-Osservatorio Astronomico di Padova, Vicolo dell'Osservatorio 5, Padova, Italy, 35122-I}
\newcommand{\ucsd}{Department of Astronomy $\&$ Astrophysics, University of California San Diego, La Jolla, CA 92093, USA}
\newcommand{\AstrobioCenter}{Astrobiology Center, 2-21-1 Osawa, Mitaka, Tokyo 181-8588, Japan}
\newcommand{\mpia}{Max Planck Institute for Astronomy, K\"onigstuhl 17, 69117 Heidelberg, Germany}
\newcommand{\stockholm}{Department of Astronomy, Stockholm University, Stockholm, SE-10691, Sweden}
\newcommand{\cambridge}{Institute of Astronomy, University of Cambridge, Madingley Road, Cambridge, CB3 0HA, UK}
\newcommand{\amnh}{Department of Astrophysics, American Museum of Natural History, Central Park West at 79th Street, NY 10024, USA}
\newcommand{\umd}{Center for Space Sciences and Technology, University of Maryland, Baltimore County, 1000 Hilltop Circle, Baltimore, MD 21250, USA}
\newcommand{\gsfcAstro}{Astrophysics Science Division, NASA-GSFC, 8800 Greenbelt Rd, Greenbelt, MD 20771, USA}
\newcommand{\gsfcCresst}{Center for Research and Exploration in Space Science and Technology, NASA-GSFC, 8800 Greenbelt Rd, Greenbelt, MD 20771, USA}
\newcommand{\CentroAstrobio}{Centro de Astrobiolog\'ia (CAB, CSIC-INTA), Camino Bajo del Castillo s/n, 28692 Villanueva de la Ca\~nada, Madrid, Spain}
\newcommand{\sfsu}{Department of Physics and Astronomy, San Francisco State University, 1600 Holloway Ave., San Francisco, CA 94132, USA}
\newcommand{\esa}{European Space Agency (ESA), ESA Office, Space Telescope Science Institute, 3700 San Martin Drive, Baltimore, MD 21218, USA}
\newcommand{\ucsc}{Department of Astronomy $\&$ Astrophysics, University of California, Santa Cruz, 1156 High Street, Santa Cruz, CA 95064}
\newcommand{\ciera}{Center for Interdisciplinary Exploration and Research in Astrophysics (CIERA), Northwestern University, 1800 Sherman Avenue, Evanston, IL 60201, USA}
\newcommand{\uvic}{Department of Physics and Astronomy University of Victoria 3800 Finnerty Road Elliott Building, Room 207. Victoria BC V8P 5C2, Canada}

\begin{document}

\title{The JWST Early Release Science Program for Direct Observations of Exoplanetary Systems IX: A Multi-Wavelength Framework for Source Characterization in JWST Coronagraphic Imaging of the HIP 65426 System}
\correspondingauthor{Evelyn L. Bruinsma}
\email{ebruins1@jhu.edu}

\author[0009-0001-0275-7811]{Evelyn L. Bruinsma}
\affiliation{\jhu}
\affiliation{\stsci}

\author[0000-0001-5365-4815]{Aarynn L. Carter}
\affiliation{\stsci}

\author[0000-0001-8074-2562]{Sasha Hinkley}
\affiliation{\exeter}

\author[0000-0003-4614-7035]{Beth A. Biller}
\affiliation{\IAedinburgh}

\author[0000-0001-6098-3924]{Andy Skemer}
\affiliation{\ucsc}

\author[0000-0001-6396-8439]{William O. Balmer}
\altaffiliation{51 Pegasi b Fellow}
\affiliation{\ciera}

\author[0009-0005-6943-6819]{Andrew D. James}
\affiliation{\exeter}

\author[0000-0002-7325-5990]{Ellis X. Bogat}
\affiliation{\marseille}

\author[0000-0002-9962-132X]{Ben J. Sutlieff}
\affiliation{\IAedinburgh}
\affiliation{\CESedinburgh}

\author[0000-0001-7739-9767]{Michael Poon}
\affiliation{\mitkavli}
\affiliation{\toronto}

\author[0000-0002-2918-8479]{Mathilde Mâlin}
\affiliation{\jhu}
\affiliation{\stsci}

\author[0000-0003-4909-256X]{Katie A. Crotts}
\affiliation{Department of Astronomy / Steward Observatory, University of Arizona, Tucson, AZ 85721, USA}

\author[0000-0002-2011-4924]{Genaro Su\'arez}
\affiliation{\amnh}

\author[0000-0003-4636-6676]{Eileen Gonzales}
\affiliation{\sfsu}

\author[0009-0008-2252-7969]{Klaus Subbotina Stephenson}
\affiliation{\uvic}

\author[0000-0001-8345-593X]{Markus Janson}
\affiliation{\stockholm}

\author[0000-0002-6510-0681]{Motohide Tamura}
\affiliation{\AstrobioCenter}

\author[0009-0000-0303-2145]{Rapha\"el Bendahan-West}
\affiliation{\exeter}

\author[0000-0002-1652-420X]{Giovanni M. Strampelli}
\affiliation{\IAedinburgh}

\author[0000-0003-1227-3084]{Michael R. Meyer}
\affiliation{\uofm}

\author[0000-0003-4203-9715]{Emily Rickman}
\affiliation{\esa}

\author[0000-0001-8627-0404]{Julien H. Girard}
\affiliation{\stsci}

\author[0000-0002-6964-8732]{Kellen Lawson}
\affiliation{\umd}
\affiliation{\gsfcAstro}
\affiliation{\gsfcCresst}

\author[0000-0002-5903-8316]{Alice Zurlo}
\affiliation{\UnivDiegoPortales}
\affiliation{\yems}

\author[0000-0001-7819-9003]{Cecilia Lazzoni}
\affiliation{\padova}

\author[0000-0002-9807-5435]{Christopher A. Theissen}
\affiliation{\ucsd}

\author[0000-0002-1493-300X]{Thomas Henning}
\affiliation{\mpia}

\author[0000-0001-7047-0874]{Pengyu Liu}
\affiliation{\IAedinburgh}
\affiliation{\CESedinburgh}

\author[0000-0001-9064-5598]{Mark C. Wyatt}
\affiliation{\cambridge}

\author[0000-0002-4388-6417]{Isabel Rebollido}
\affiliation{\CentroAstrobio}

\author[0009-0004-0868-1186]{Aiza Kenzhebekova}
\affiliation{\IAedinburgh}
\affiliation{\CESedinburgh}

\begin{abstract}
We present an updated analysis of the JWST coronagraphic imaging data of the HIP 65426 system. We aim to search for and characterize potential planetary companions as well as provide updated astrometry and photometry for the known exoplanet HIP~65426~b following updates to the JWST and spaceKLIP pipelines. Our measured photometry and astrometry show significant differences to those in \cite{carter2023} but are consistent to within $2.2\sigma$ in separation, $1.5\sigma$ in position angle, and $3.7\sigma$ in flux. Our results indicate that pipeline updates have changed the photometry and astrometry significantly in some filters, especially in terms of observed flux. We see the greatest changes in separation in F444W and F1550C, position angle in F250M, and flux in F250M, F300M, F410M, F1140C, and F1550C. The SED of HIP~65426~b remains consistent with a young, giant planet best fit by models with a cloudy atmosphere and an effective temperature estimate of $\sim$1436 K. Using both NIRCam and MIRI data, we identified 21 candidate companion sources in addition to the known planet. Of the 21 sources, 13 were attributed to astrophysical sources with eight background stars, four galaxies, one unknown, and zero planets. While one of these sources was detected in \cite{chauvin2017}, the remaining point sources were undetected in previous ground-based observations, which makes this work an initial epoch of astrometry for these objects. One galaxy in particular appears as a point source in MIRI but is highly extended in NIRCam and exposes a potential drawback to single-epoch MIRI imaging in a blind search for companions. 
\end{abstract}

\section{Introduction} \label{sec:intro}
High-contrast imaging is a key technique used to observe faint targets, such as exoplanets or disks, in close proximity to their bright host stars. JWST Early Release Science Program 01386 \citep{hinkley2022a} was designed to test and demonstrate JWST's capabilities for both high-contrast coronagraphic imaging and spectroscopic observations of widely-separated exoplanet companions. This program is now complete and included coronagraphic imaging of HIP~65426~b \citep{carter2023} and the circumstellar disk of HD 141569 A \citep{millar2025} from 2-16~\um, spectroscopy from 1-28~\um~of the planetary mass companion VHS J125601.92-125723.9 AB b \citep{miles2023, petrus2024, mathilde, vhs1256_deRegt}, and Aperture Masking Interferometry observations of HIP 65426 at 3.8~\um~ \citep{sallum2024, ray2025}. 

The main goals of this program were to create datasets for exoplanet and disk imaging users of commonly used modes, provide science enabling products to the community to assist with future proposals and investigations, characterize exoplanets over their full spectral range (2-28~\um) for the first time, and spectrophotometrically characterize disks out to 15~\um~ \citep{ers}. 

HIP~65426~b was characterized in detail in \cite{carter2023}, which revealed the first ever detection of an exoplanet beyond 5~\um; highlighted JWST's higher than expected contrast performance, which is sensitive to companions with masses as low as 0.3~\mjup{} at $\sim$100~AU in the case of HIP~65426~b; obtained very precise measurements of HIP~65426~b's bolometric luminosity, $\text{L}_{bol}/$\lsun$=-4.14$ to $-4.31$; and provided high-precision photometric measurements of the planet across seven filters from $1-16$~\um.  

In this work, we revisit the JWST coronagraphic imaging observations of HIP~65426, as \cite{carter2023} focused only on the known planet HIP~65426~b without considering the background sources visible in the images. We aim to analyze and characterize these additional sources, provide an epoch of astrometry for these sources to aid in proper motion checks in future observations, and verify whether the photometry and astrometry from \cite{carter2023} still hold with the updated data reduction pipeline.

HIP~65426~b is a $7.1\pm1.2$~\mjup{} planet with a semi-major axis of $86^{+116}_{-31}$~AU\footnote{\cite{carter2023} states an incorrect value of $87^{+108}_{-31}$~AU in the abstract with the correct value of $86^{+116}_{-31}$~AU in \S~5.1} \citep{carter2023} and is located at a distance of $107.49\pm{0.40}$~pc \citep{gaia2021}. While disk instability could explain the planet's current location \citep{grav_instability}, this wide separation orbit is challenging to form via core accretion \citep{accretion}. For the planet to have formed via core accretion, formation would have occurred significantly closer to the star followed by a planet-planet scattering event, which would require closer-in giant planets early on in the system \citep{scattering}. If this is indeed the case, there remains the possibility that there exist additional wide-separation planets that were scattered in a similar manner. Identifying the presence or absence of additional planets could provide valuable insights into the dynamics of this system and help explain how HIP~65426~b ended up at its current location. JWST observations provide a good opportunity to search for additional planets, as they have the greatest sensitivity to lower-mass companions compared to ground-based observations \citep{carter2021, twa7b_lagrange, twa7b_katie}. 

In the last three years, the JWST pipeline and its calibration reference files have evolved and been updated compared to those available upon commissioning. Furthermore, the \texttt{spaceKLIP} reduction pipeline \citep{kammerer2022, carter2023} and its supporting packages, such as \texttt{STPSF} \citep{perrin2012, perrin2014}, have also seen updates and changes during that time. \texttt{spaceKLIP} is a publicly available \texttt{python} package that generates point spread function (PSF) subtracted images and contrast curves as well as photometry and astrometry of associated companions using products from the JWST pipeline and various functionalities from the \texttt{pyKLIP} \citep{pyklip} package. While the main objective of this work is to investigate the nature of the other sources in the images, we will also conduct a reanalysis of the photometry and astrometry of HIP~65426~b with the updated pipeline. Comparing the values obtained in this work to those from \cite{carter2023} will provide insight into any significant changes due to updates in both the JWST and \texttt{spaceKLIP} pipelines.

In Section~\ref{sec:obs}, we provide a description of the JWST observations and the data reduction process as well as cover the methods used to identify, extract, and vet the candidate companion (CC) sources based on their photometry. We provide a discussion and analysis of the results in Section~\ref{sec:disc}, and we summarize our conclusions in Section~\ref{sec:conclusion}.

\section{Observations and Data Reduction} \label{sec:obs}
\subsection{Observations}
HIP 65426 was observed with the Mid-Infrared Instrument (MIRI) on July 17, 2022 and with the Near Infrared Camera (NIRCam) on July 29-30, 2022. All observations of HIP~65426~b consist of two science images repeated at two independent roll angles separated by $\sim10^\circ$. A bright reference star, HIP 68245 (B2IV, 2MASS J=4.628~mag, J-K=0.137~mag) was also observed with a nine-point dither pattern to allow for PSF subtraction. The NIRCam images were taken in each of the F250M, F300M, F356W, F410M, and F444W filters. The NIRCam reference observations used the 9-POINT-CIRCLE dither pattern. The MIRI observations were taken in the F1140C and F1550C filters, and the reference images utilized the 9-POINT-SMALL-GRID dither pattern. The MIRI observations also included four background images in both filters to address the stray-light ``glowstick" effect that is present in MIRI coronagraphic images \citep{boccaletti2022}. NIRCam short wavelength measurements were not taken, as dual-band imaging in the NIRCam coronagraphic mode was not available until June 2023. 

\pagebreak
\subsection{Data Reduction}
\label{subsec: data reduction}
The data were reduced using the \texttt{spaceKLIP}\footnote{v2.2.1.dev70+g6be81e4dc, \url{https://github.com/spacetelescope/spaceKLIP}} and JWST\footnote{All data were processed using pipeline version CAL\_VER=1.20.2 and calibration reference data CRDS\_CTX=jwst\_1464.pmap} pipelines. For the NIRCam data, Stage 0 (\textit{*uncal.fits}) were processed into Stage 1 (\textit{*rateints}) products using spaceKLIP's implementation of the JWST pipeline. This stage performs detector-level corrections, including group-by-group corrections and ramp fitting. For the ``jump" parameter, which flags jumps in the ramp between two groups that are large relative to other groups, the default value of 4 was used. One exception to this was for F410M, which used a value of 1 to help mitigate the effects of a very spatially large cosmic ray. For this data, we skipped the dark current subtraction and included a 1/f noise correction to eliminate instrumental striping, as described in \cite{rebollido2024}. The Stage 1 products were then processed to Stage 2 (\textit{*calints.fits}) products by performing instrument-level corrections, such as flat-fielding and photometric calibration. Lastly, the corrected images were prepared for post-processing with bad pixel correction using data quality arrays, sigma clipping, and 2D interpolation. We first ran the \texttt{find\_bad\_pixels} function with the ``dqarr+sigclip" method. The sigclip parameters included a sigma clipping threshold of five, a sigma clipping threshold for negative outliers of five, and a shift of -3 to 3 in both the x- and y-directions. The pixels were then cleaned using the \texttt{clean\_bad\_pixels} function using the ``interp2d" method with a kernel size of 9. Due to JWST's known undersampling at shorter wavelengths, the F250M images were blurred with a Gaussian filter in order to achieve the Nyquist sampling necessary for the image shifting. Following this, we centered and aligned the frames as well as performed a NIRCam specific background subtraction as described in \cite{lawson2024}. \cite{carter2023} performed a median subtraction in place of this NIRCam specific background subtraction, as this method did not yet exist. Lastly, the frames were padded with 80 pixels in preparation for PSF subtraction. 

The MIRI reduction followed a similar procedure as the NIRCam images with minor adjustments. During the Stage 1 processing, a jump threshold of 8 was used as lower values resulted in an unrealistic number of pixels being flagged as having a jump. The flat-field correction was skipped, as it increases noise and the ``glowstick" effect near the mask center \citep{boccaletti2024}. After the Stage 2 processing, the MIRI images were cropped to remove regions outside of the coronagraphic field of view. The MIRI background is spatially correlated due to the ``glowstick" structure described in \cite{boccaletti2022}, and the detector has a spatially structured, temporally varying noise known as reset switch charge decay. Therefore, it is recommended to use the background subtraction method for MIRI as described in \cite{godoy2024} rather than a median subtraction, as this method allows for both of these effects to be captured. This step also differs from \cite{carter2023}, as this method did not exist at the time.

Following calibration and cleaning, PSF subtraction was performed using \texttt{pyKLIP}\footnote{v2.9.1, \url{https://pyklip.readthedocs.io}}, which applies a principal component analysis (PCA) framework. This was achieved using a combination of the angular differential imaging (ADI) \citep{marois2006} and reference differential imaging (RDI) \citep{soummer2012} methods using one annulus, one subsection, and a range of KL-modes from 1-40. This step addresses any residual stellar flux that remains despite the use of a coronagraph and allows us to achieve a higher sensitivity to faint companions at the closest separations.

\begin{figure*}
    \centering
    \includegraphics[width=0.8\linewidth]{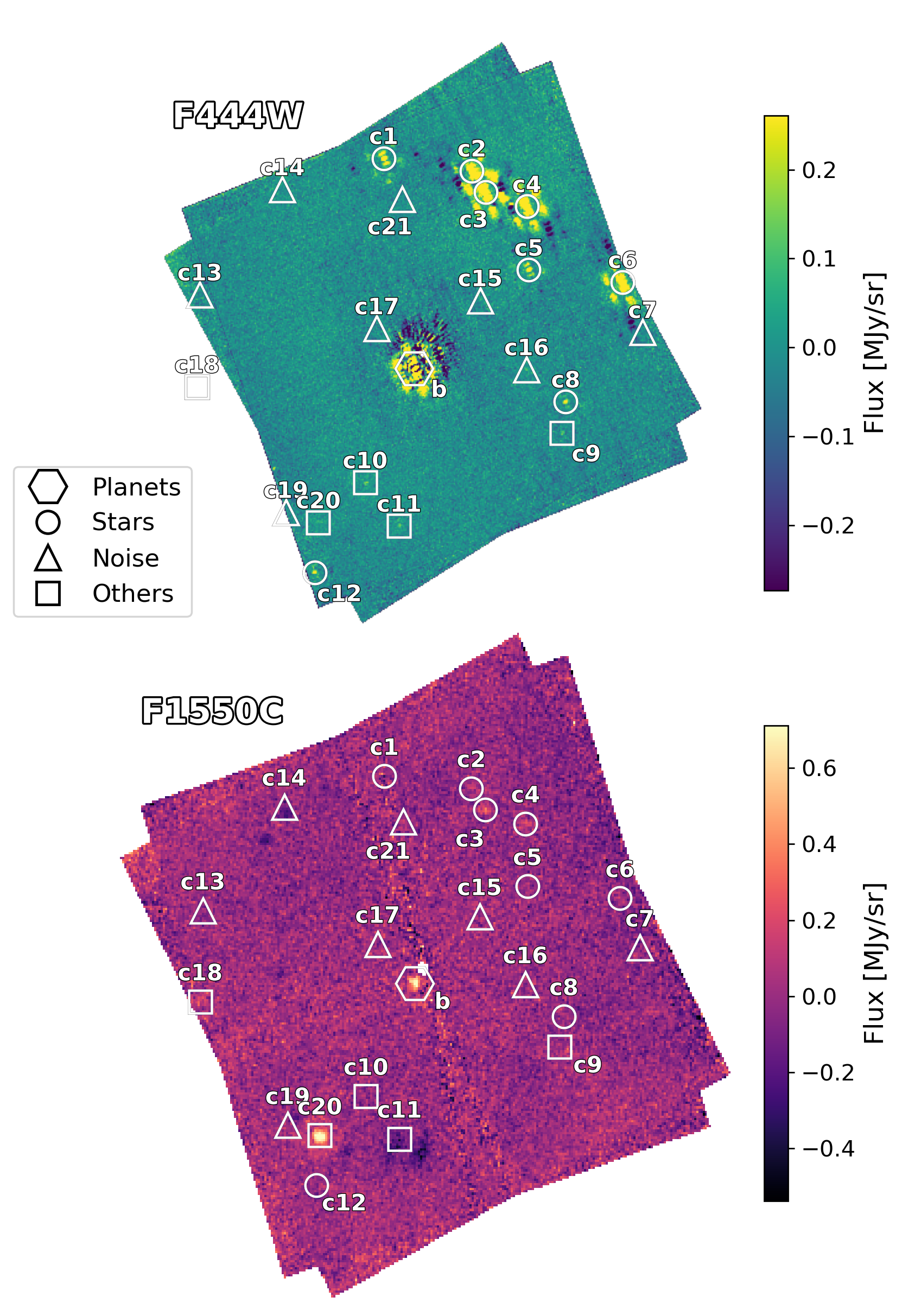}
    \caption{Subtracted images for the NIRCam F444W (top) and MIRI F1550C (bottom) filters. The images are oriented with North pointing up and East pointing left. All candidate companions pre-vetting are marked with white markers. Not all sources were detected in all filters, so some markers appear empty. }
    \label{fig:sources_f444_f1550}
\end{figure*}

\subsection{Source Fitting}
Between the NIRCam and MIRI images, 21 potential CCs were identified by visual inspection. Figure~\ref{fig:sources_f444_f1550} shows these sources marked in two example filters (F444W and F1550C) with the remaining filters shown in Figure~\ref{fig:sources_others}. Some sources were detected in only one or two filters, so some markers appear empty, and there are some negative lobes in the images caused by the ADI roll. For the MIRI images, there are additional negative sources that overlap with c11 and c14 which were introduced from the background images. Previous attempts have been made at using an automated source detection algorithm, but the results were inconsistent with some sources being detected multiple times or being skipped over entirely (GO 4050, GO 5835, SURVEY 6005, priv. comm.). The complex spatial structure of the NIRCam PSF complicates and inhibits traditional automatic detection methods and is distinct from ground-based scenarios where the PSF is close to a gaussian. An automated routine that encapsulates the JWST PSF structure may be able to discern and discriminate sources more clearly than a visual inspection; however, that is beyond the scope of this work. 

The best-fit parameters for these CCs were determined using the \texttt{extract\_companions} function in \texttt{spaceKLIP}. This process takes RA and DEC offsets of their approximate locations using visually identified pixel positions, the estimated contrast of each companion, and a stellar model of the host star, which was obtained from the publicly available \texttt{spaceKLIP} tutorials and is identical to the model used in \cite{carter2023}. A Markov Chain Monte Carlo (MCMC) framework was then used to derive the best-fit position and photometry of each CC with 100 walkers for 800 steps with a burn-in of 800 steps. The PSF models for each filter were obtained from \texttt{webbpsf\_ext}\footnote{v2.0.1.dev529, \url{https://github.com/JarronL/webbpsf_ext}}. For consistency with the original analysis, the default diagonal kernel was used for the fit. 

For the NIRCam photometric errors, we used the method of injection and recovery described in \cite{carter2023}. For each source, we subtracted the companion flux as measured by the MCMC process, injected a source of the same contrast (generated with \texttt{STPSF}\footnote{v2.2.0, \url{https://stpsf.readthedocs.io/en/latest/}}) at the same separation at ten equally-spaced angles, being careful not to overlap with any subtracted sources. Each injection was placed in its own image rather than injecting them all into the same image. Ten separation angles were chosen because any number greater than this significantly increased the computational time but had no effect on the calculated error. The same PSF fitting procedure described above was then implemented on these synthetic PSFs, and the standard deviation of these flux measurements were combined with the absolute calibration percent error (5\% for NIRCam\footnote{see Table 1, \url{https://jwst-docs.stsci.edu/jwst-calibration-status/jwst-absolute-flux-calibration\#gsc.tab=0}\label{fn:flux_cal}}) and the stellar photometric error measured by \texttt{spaceKLIP} where the standard deviation was the dominant term in the uncertainty calculation. For the MIRI filters, we did not see evidence for speckle noise or correlated noise, so this method of injection and recovery was deemed unnecessary, and we adopted the errors 
generated by \texttt{spaceKLIP} combined with the absolute calibration percent error (15\% for MIRI\footref{fn:flux_cal}) and the stellar photometric error measured by \texttt{spaceKLIP}, where the dominant term was the \texttt{spaceKLIP} calculated error.

\subsection{Candidate Vetting} \label{sec:sources}
After fitting the PSF of each CC across all wavelengths, we conducted a vetting process to determine if a given detected source is astrophysical or an image artifact. Through this vetting process, we split the sources into three categories: point sources, extended sources, and a general discard category. This was achieved with a manual, visual inspection of the companion fit for each source in each filter.

We started with the discard category. For each source, the observed signal was compared to the best-fit model PSF for consistency. We also looked at repeatability across multiple filters. If the source morphology was inconsistent with the model PSF it was sorted into the discard category. Such examples included cosmic rays and bad pixels. Additionally, if the source was only visible in one filter, it was also thrown out; however, sources only visible in the MIRI wavelengths were still considered, as these wavelengths are more sensitive to low-mass planets. At distances beyond 60~pc, NIRCam F444W becomes more advantageous than MIRI for planet detection, as detectable planets are generally hotter \citep{nircam_yells_at_cloud}. Given that HIP 65426 is farther than 60~pc, we also consider sources visible only in F410M and F444W.

\begin{figure}
  \centering
  \includegraphics[width=\linewidth]{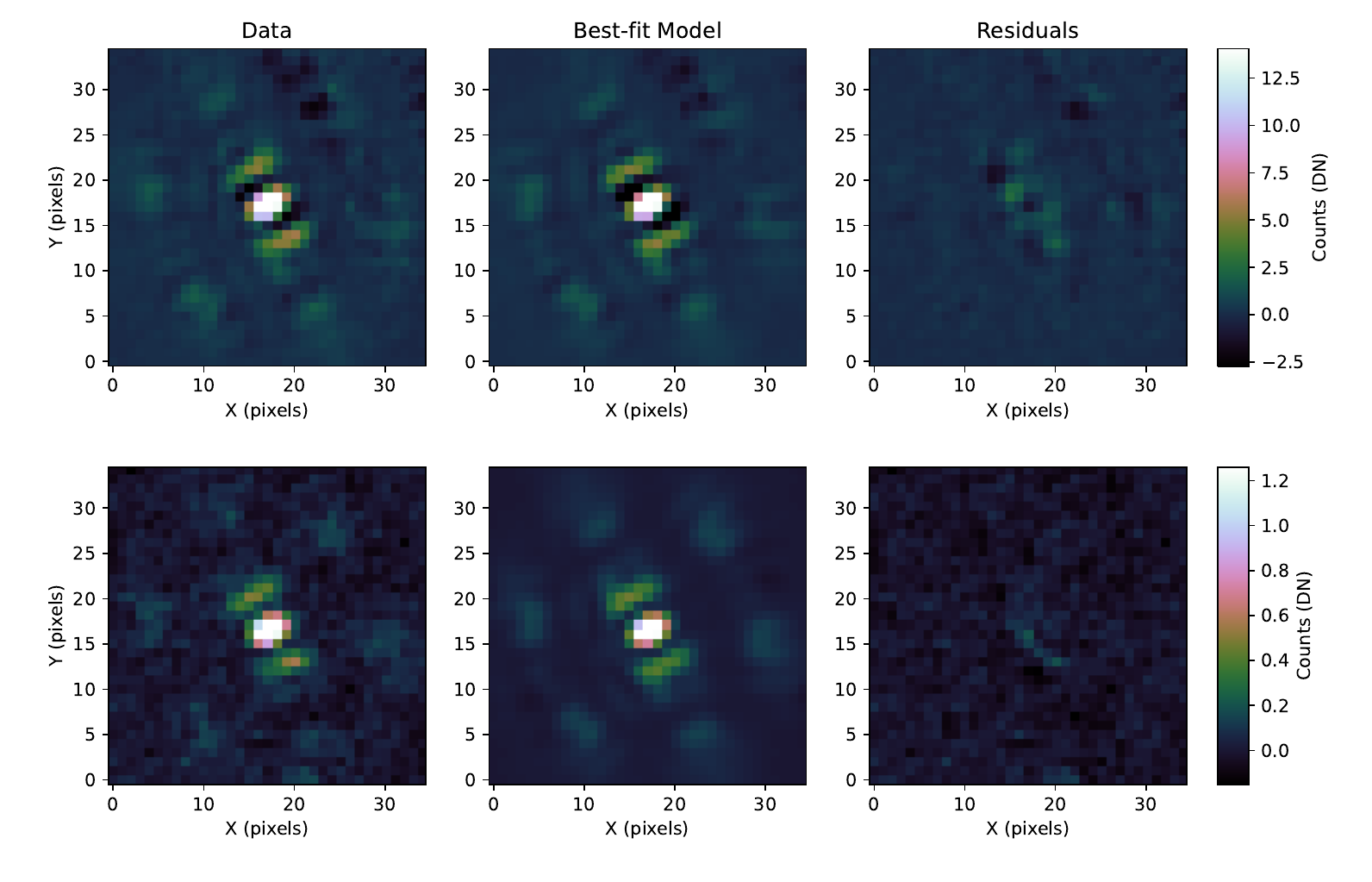}
  \caption{Point source companion fit example in F444W for HIP~65426~b (top) and c1 (bottom).} 
  \label{fig:fit_example}
\end{figure}

After removing the contaminants, we split the remaining sources into the point source and extended source categories based on their correlation with the model PSF. Point sources will match the best-fit model almost exactly with very few residuals, as shown for c1 in Figure~\ref{fig:fit_example}. Extended sources, on the other hand, have a spatially extended structure that is inconsistent with a point source PSF. Of the remaining sources, all were deemed to be point sources with one exception: c20 appears point-like in the mid-infrared (mid-IR), while appearing extended in F356W and F444W with non-detections in the remaining NIRCam filters, as shown in Figure~\ref{fig:psf_c18c20}. Even though it is slightly visible in F356W and F444W, we were unable to recover statistically significant flux values, so these were treated as non-detections. For the non-detections, the measured flux was treated as being consistent with zero with $3\sigma$ upper limits. 

c9, c10, and c11 appeared visually as very faint point sources but were unable to be properly fit by the PSF models. They consistently failed to capture an accurate shape and converged at too low of a flux, even after varying the \texttt{frange} values in \texttt{extract\_companions}. They were then refit using a Matérn 3/2 kernel with a Powell minimization method. Described in \cite{lazzoni2025}, the Powell minimization method is a routine that blurs the PSF before fitting. By using a Gaussian convolved model PSF, we were able to more accurately model the shape and brightness of these sources and capture the spatially correlated noise across nearby pixels. This method is more effective when fitting extended sources, so this was our first clue that these objects might be galaxies. For consistency with all of the extended sources, this was also applied to c20.

The PSF for c8 was also not properly fit by the PSF models, and by altering the parameters in the pixel cleaning routine and the \texttt{extract\_companions} function, we obtained a more accurate fit. For the pixel cleaning routine, we increased the sigma clipping threshold to six, the sigma clipping threshold for negative outliers to six, and the interp2d kernel size to ten). For the \texttt{extract\_companions} function, we set the \texttt{frange} parameter to 0.75, which restricted the flux range to be within the bounds $1/10^{0.75}$ to $10^{0.75}$ times the initial guess flux, as the fit was converging at too low of a flux. Additionally, we employed a Matérn 3/2 kernel, and the Powell minimization method offered no improvement, which suggests the existence of non-Gaussian noise near the PSF, such as a bad pixel aligned with the PSF core. While it would not be reliably disentangled from the PSF structure during the pixel cleaning process, the Matérn 3/2 kernel would be able to mitigate its effects.

Out of the 21 original sources detected in the images, 13 survived the vetting process. The PSF fits for each vetted source in each filter, in addition to that of HIP~65426~b, are shown in Appendix~\ref{sec:companion_fits}. A summary table of the vetted background sources and their associated parameters is also shown in Table~\ref{tab:sources_table} in the appendix, whereas HIP~65426~b and its associated parameters are shown in Table~\ref{tab:table_b}.

\begin{table*}
    \centering
    \caption{JWST astrometry and photometry of HIP~65426~b. The position angle ($\theta$) is provided from North through East. All apparent magnitudes are relative to Vega. The uncertainties in astrometry are propagated with an additional uncertainty of 0.1 pixels (6.3~mas for the NIRCam filters and 10~mas for the MIRI filters) to account for the assumed precision of star centering.}
    \label{tab:table_b}
    \begin{tabular}{lllllll}
\toprule
Source & Filter & $\rho$ (mas) & $\theta$ (deg) & Flux (Jy) & Flux (Wm$^{-2}$\um$^{-1}$) & m$_b$ (mag) \\
\midrule
b & F250M & 817 $\pm$ 7 & 150.7 $\pm$ 0.46 & (1.17 $\pm$ 0.12) $\times$ 10$^{-4}$ & (5.58 $\pm$ 0.56) $\times$ 10$^{-17}$ & 16.590 $\pm$ 0.103 \\
 & F300M & 814 $\pm$ 6 & 149.6 $\pm$ 0.45 & (1.25 $\pm$ 0.09) $\times$ 10$^{-4}$ & (4.16 $\pm$ 0.28) $\times$ 10$^{-17}$ & 16.172 $\pm$ 0.084 \\
 & F356W & 816 $\pm$ 6 & 149.5 $\pm$ 0.44 & (1.52 $\pm$ 0.10) $\times$ 10$^{-4}$ & (3.58 $\pm$ 0.22) $\times$ 10$^{-17}$ & 15.626 $\pm$ 0.075 \\
 & F410M & 812 $\pm$ 6 & 149.5 $\pm$ 0.45 & (1.55 $\pm$ 0.10) $\times$ 10$^{-4}$ & (2.80 $\pm$ 0.18) $\times$ 10$^{-17}$ & 15.330 $\pm$ 0.077 \\
 & F444W & 809 $\pm$ 6 & 150.0 $\pm$ 0.45 & (1.39 $\pm$ 0.09) $\times$ 10$^{-4}$ & (2.19 $\pm$ 0.15) $\times$ 10$^{-17}$ & 15.322 $\pm$ 0.080 \\
 & F1140C & 829 $\pm$ 12 & 148.0 $\pm$ 0.80 & (4.95 $\pm$ 0.79) $\times$ 10$^{-5}$ & (1.16 $\pm$ 0.19) $\times$ 10$^{-18}$ & 14.444 $\pm$ 0.174 \\
 & F1550C & 786 $\pm$ 17 & 150.7 $\pm$ 1.21 & (3.67 $\pm$ 0.67) $\times$ 10$^{-5}$ & (4.57 $\pm$ 0.84) $\times$ 10$^{-19}$ & 14.096 $\pm$ 0.198 \\
\bottomrule
\end{tabular}
   
\end{table*}

\section{Discussion} \label{sec:disc}
\subsection{Color Magnitude Diagrams}
\begin{figure*}
    \centering  {\includegraphics[width=0.8\textwidth]{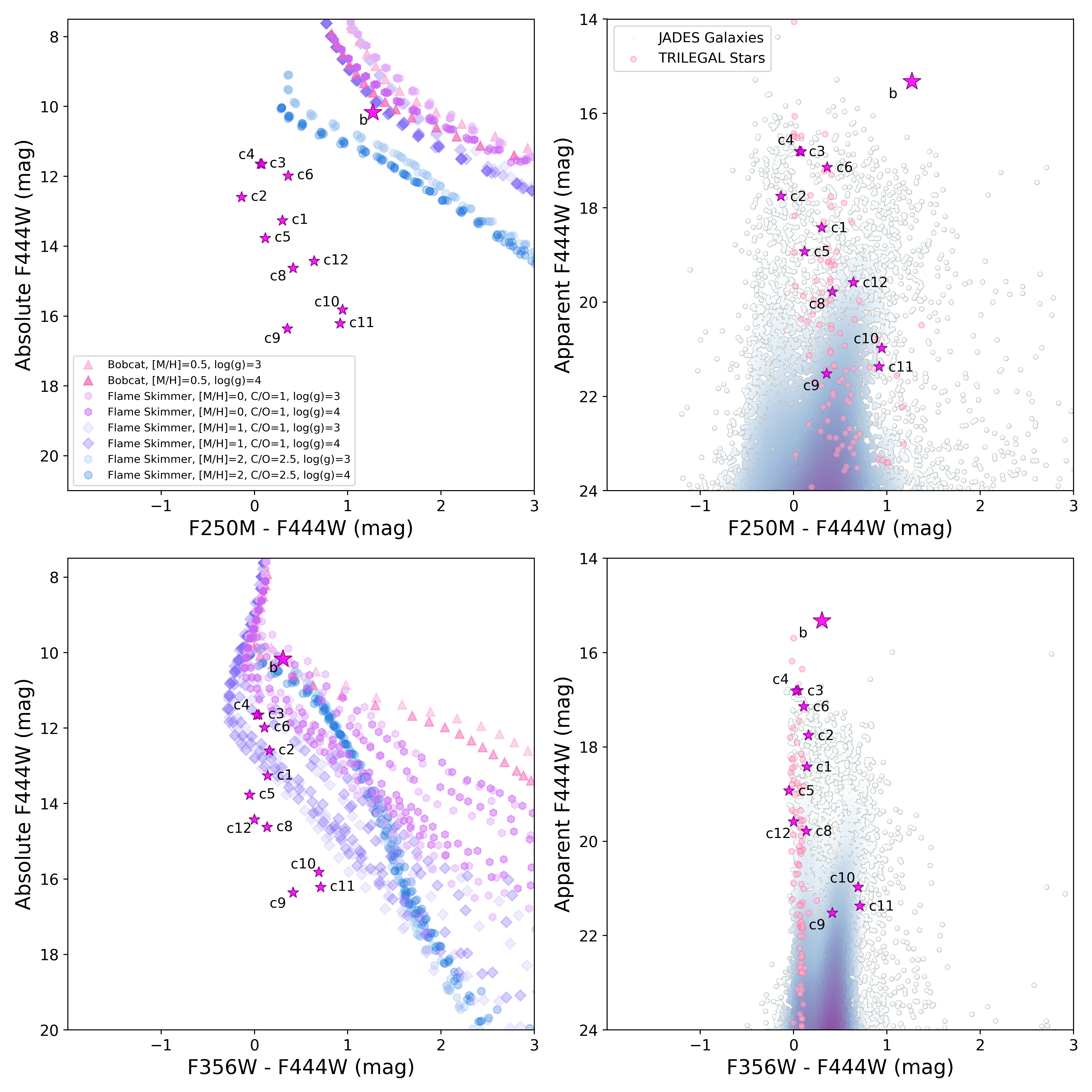}} 
  \caption{Color magnitude diagram of all vetted companion candidates. \textbf{Left:}  Absolute magnitude in F444W vs F250M-F444W (\textbf{top}) and F356W-F444W (\textbf{bottom}). The candidates are plotted alongside Sonora Bobcat and Sonora Flame Skimmer of varying metallicity, log(g), and C/O. In both cases, HIP~65426~b falls directly on the planetary model tracks. For F356W-F444W, the remaining sources overlap with the planetary atmosphere tracks, whereas for F250M-F444W, the remaining sources are too faint to intersect with the tracks. 
  \textbf{Right:} Apparent magnitude in F444W vs F250M-F444W (\textbf{top}) and F356W-F444W (\textbf{bottom}). The candidates are plotted alongside simulated TRILEGAL stars in the direction of HIP 65426 as well as galaxies from the JADES survey. For F250M-F444W, there is no clear distincion between the extragalactic and stellar candidate sources, whereas for F356W-F444W, eight of the 13 sources align with the TRILEGAL star colors, and HIP~65426~b hovers above. The remaining three sources appear redder than the rest and intersect with the JADES population, indicating that these sources are background galaxies. c18 and c20 are not included, as there is no available NIRCam data for c18 and there are only NIRCam upper limits available for c20.}
  \label{fig:cmd}
\end{figure*}

To assess whether any of the detected point sources are consistent with substellar companions, we considered the absolute F444W magnitude vs. F250M-F444W and F356W-F444W colors, as shown in Figure~\ref{fig:cmd} (left panel). Also plotted are two suites of models evaluated at the assumed age of the star ($\sim$14 Myr \citep{gagne2018}): Sonora Bobcat models \citep{sonora_bobcat} for two values of \logg{}, and the Sonora Flame Skimmer models \citep{flame_skimmer_paper} for various metallicities, C/O, and \logg. Assuming the HIP 65426 distance of $107.49\pm 0.40$ pc \citep{gaia2021} for the background sources, any planets would fall on or near these model tracks. Due to sensitivities to assumptions about metallicity and vertical mixing, we were unable to distinguish between any potential stellar or planetary objects in F356W-F444W; however, in F250M-F444W, every source apart from HIP~65426~b is too faint or too blue to be consistent with planetary-mass objects bound to the host star for metallicities from 0 to 2 dex and C/O ratios from 1 to 2.5.

We also considered the apparent F444W magnitude vs. F250M-F444W and F356W-F444W colors in which we plotted simulated stars from a TRILEGAL galactic population synthesis model\footnote{\url{https://stev.oapd.inaf.it/cgi-bin/trilegal}} \citep{trilegal} and galaxies from the JADES survey \citep{jades_catalog, jades_overview} (Figure~\ref{fig:cmd}, right panel). We ran the TRILEGAL population synthesis model with the default parameters at the coordinates of HIP 65426. We plotted 100 random stars from the TRILEGAL model as an example of what the distribution of background stars would look like along the line of sight of HIP 65426. For F250M-F444W, the TRILEGAL stars were too spread out in color to yield a definitive separation between the extragalactic and stellar candidates, but in F356W-F444W, eight of the CCs overlap with the TRILEGAL stellar population, which indicates that they are likely stellar in nature. The remaining three sources (c9, c10, and c11) are significantly redder and overlap with the JADES galaxies, which indicates they are likely extragalactic in nature. c18 and c20 were not included, as there is no available NIRCam data for c18 and there are only NIRCam upper-limits available for c20. 

\subsection{SED Analysis}
\label{subsec: SED Analysis}
As an additional check to confirm the findings from the color magnitude diagram, we characterized these sources by plotting their individual spectral energy distributions (SEDs) using the fluxes obtained during the companion fitting process. For each source, we fit the SEDs with a variety of atmospheric models using \texttt{species}\footnote{v0.10.4, \url{https://species.readthedocs.io/en/latest/}} \citep{species} to determine the best-fit model for each source. Such models included Phoenix \citep{phoenix_husser2013}, BT-Settl \citep{bt_settl1, bt_settl2}, ATMO \citep{atmo, atmo_petrus2023}, AMES \citep{ames_cond}, Exo-REM \citep{exorem}, Sonora Bobcat \citep{sonora_bobcat}, and Sonora Diamondback \citep{sonora_diamondback}. Using the flux density at each observed wavelength for each source as well as a parallax for HIP 65426 b or a range of parallaxes for the CCs (0.1 to 100 mas) and a range of effective temperatures (the full coverage range for each model) over which the model grids are generated, a Bayesian inference was made using nested sampling \citep{pymultinest, multinest1, multinest2, multinest3} in order to fit the data to the chosen model grid.

\begin{figure*}
  \centering
  \includegraphics[width=0.85\linewidth]{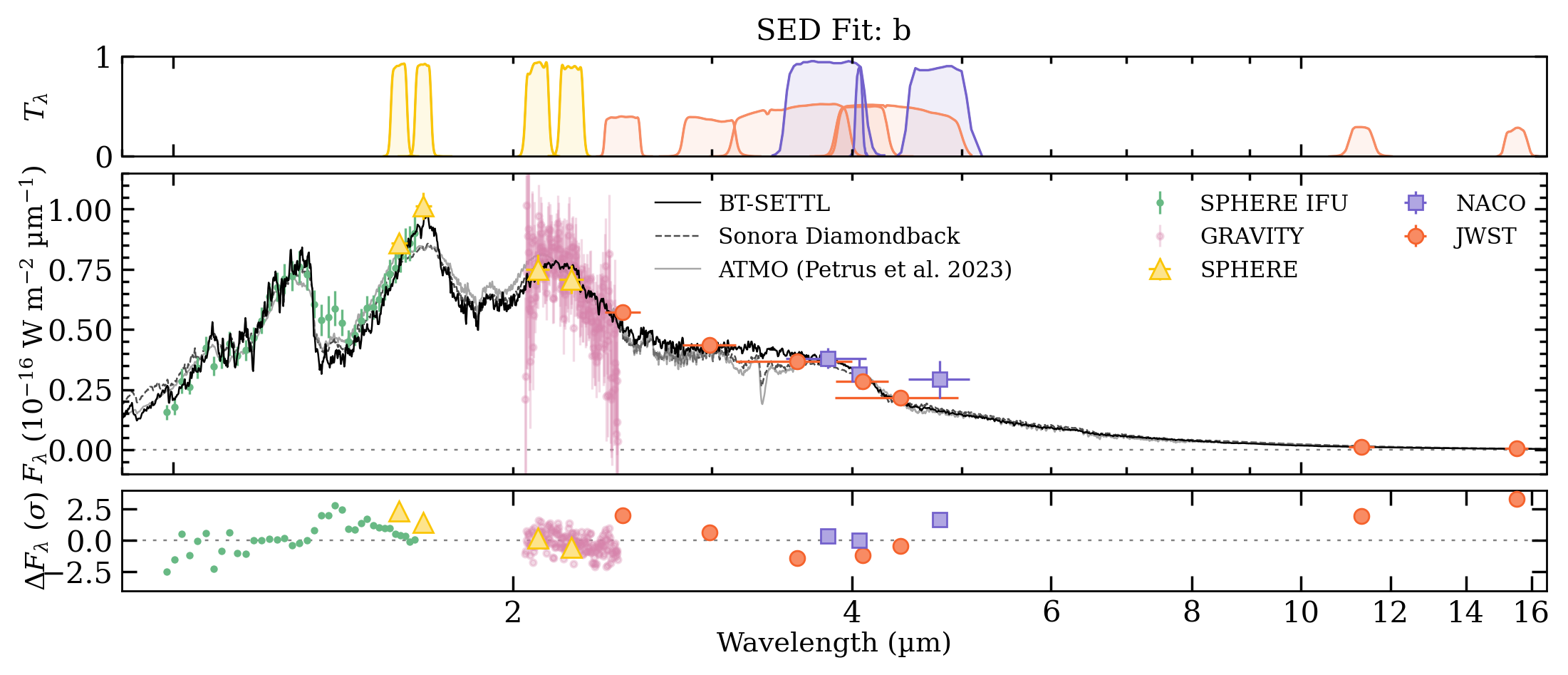}
  \includegraphics[width=0.85\linewidth]{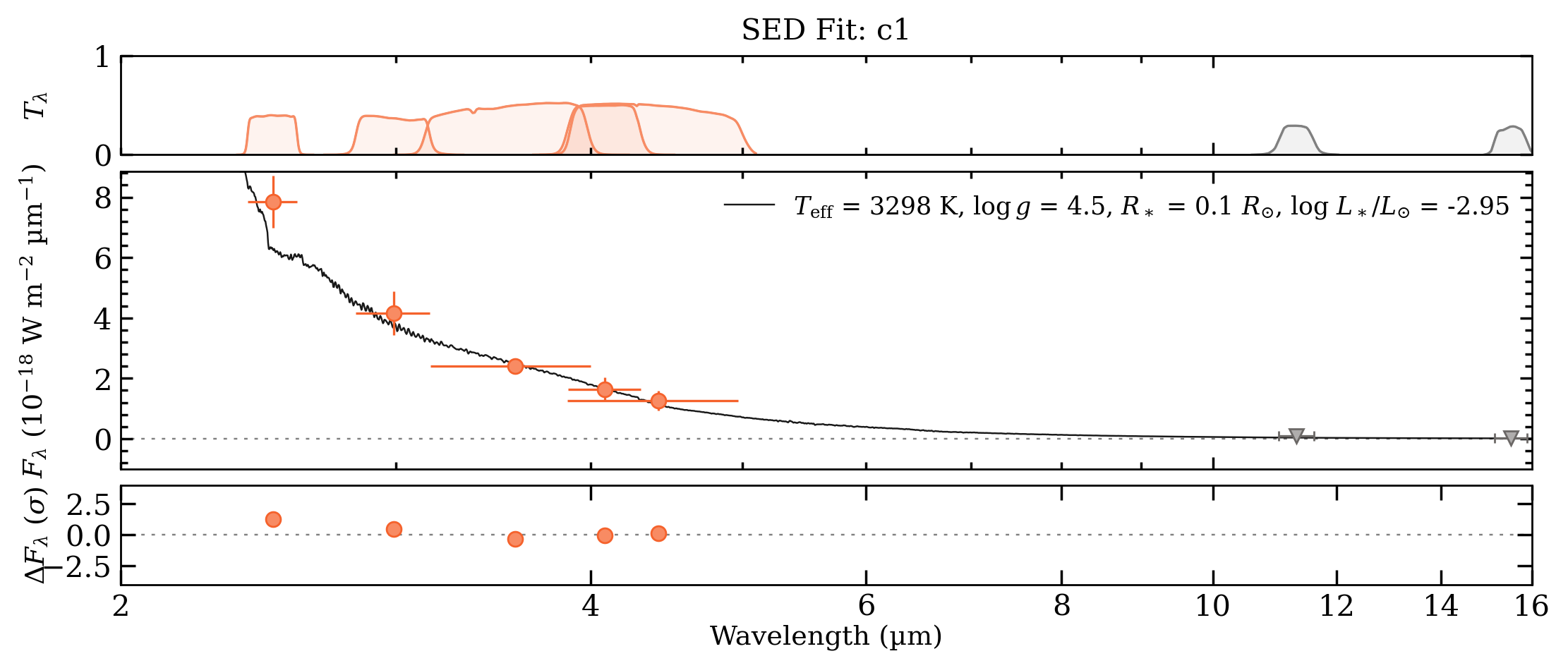}
  \includegraphics[width=0.85\linewidth]{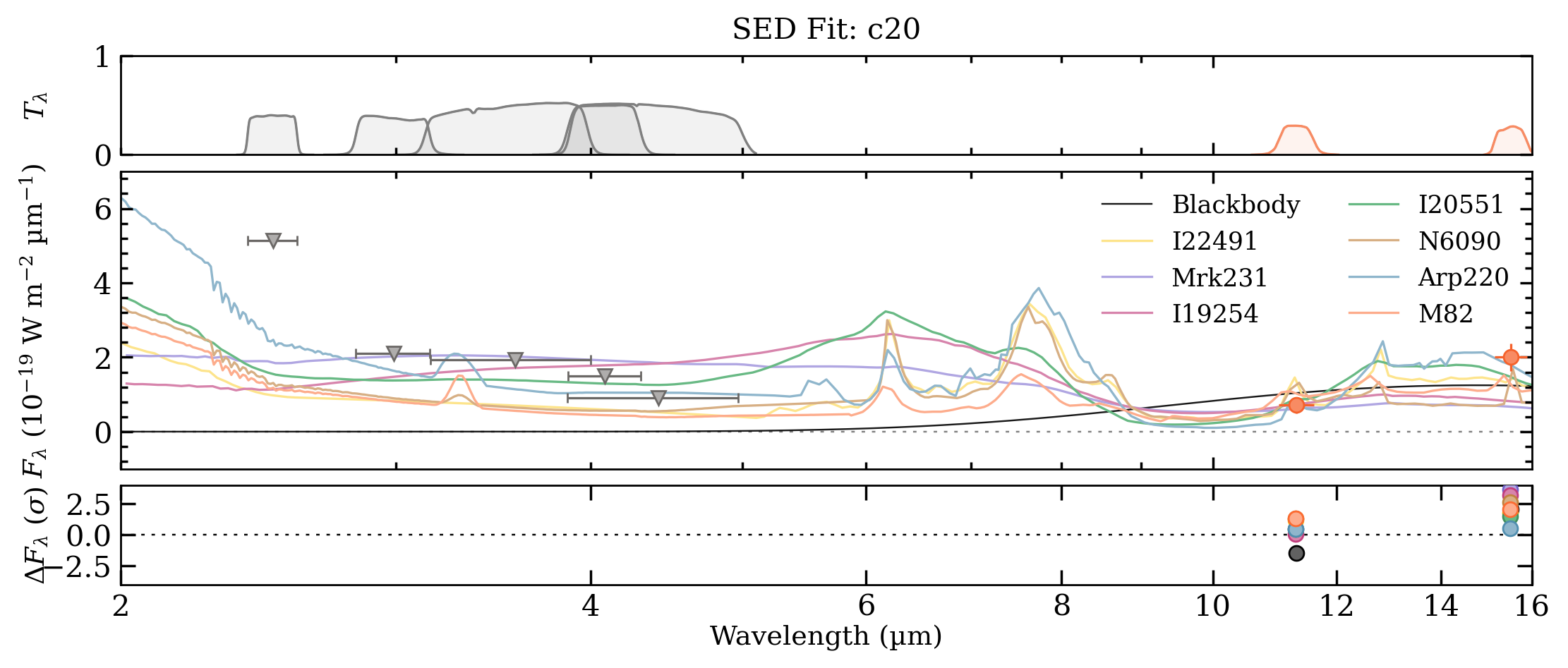}
  \caption{Example SED fits for each type of object. For each panel, the top is the bandpasses of the photometric data points, the middle is the data plotted alongside the best-fit models, and the bottom is the residuals of each data point relative to the best-fit model. $3\sigma$ upper-limits are shown in grey. 
  \textbf{Top: }SED of HIP~65426~b, using data from JWST, VLT/SPHERE \citep{chauvin2017}, VLTI/GRAVITY \citep{blunt2023}, and VLT/NACO \citep{cheetham2019}, plotted alongside the best-fit BT-SETTL (\teff{} $=1436^{+13}_{-12}$~K, \logg{} $=3.64\pm0.02$~dex, and $\text{R}_\text{p}=1.43\pm{0.03}$~\rjup), Sonora Diamondback (\teff{} $=1210^{+9}_{-7}$~K, \logg{} $=3.76\pm0.06$~dex, and $\text{R}_\text{p}=1.86\pm{0.03}$~\rjup), and ATMO \citep{atmo_petrus2023} (\teff{} $=1376\pm14$~K, \logg{} $=3.87^{+0.07}_{-0.08}$~dex, and $\text{R}_\text{p}=1.50\pm{0.03}$~\rjup).
  \textbf{Middle: }Example SED of a background star (c1) plotted alongside the best-fit BT-SETTL model, which yields \teff{} $=3298^{+447}_{-572}$~K, \logg{} $=4.5^{+0.64}_{-0.62}$~dex, and $\text{R}=0.10^{+0.007}_{-0.008}$~\rsun. 
  \textbf{Bottom: }Example SED of a background galaxy (c20) plotted alongside the best-fit blackbody and galaxy templates. }
  \label{fig:seds_main}
\end{figure*}

\subsubsection{HIP~65426~b}
\label{subsubsec: HIP~65426~b}
The SED of HIP~65426~b, shown in Figure~\ref{fig:seds_main}, remains consistent with the SED of a young, giant planet. We included archival data from VLT \citep{chauvin2017, cheetham2019, blunt2023} to provide additional context for this object. The SED was fit with various atmospheric models including BT-SETTL, Sonora Diamondback and ATMO \citep{atmo_petrus2023}. 

From the BT-SETTL model, we find that HIP~65426~b's flux distribution is described by an effective temperature of $1436^{+13}_{-12}$~K with a cloudy atmosphere and a \logg{} of $=3.64\pm0.02$~dex, compared to the \teff$=1624_{-15}^{+16}$ K, \logg$=3.88\pm{0.08}$~dex reported in \cite{carter2023}. The Sonora Diamondback fit yields \teff{} $=1210^{+9}_{-7}$~K, \logg{} $=3.76\pm0.06$~dex, and $\text{R}_\text{p}=1.86\pm{+0.03}$~\rjup, and the ATMO fit yields \teff{} $=1376\pm14$~K, \logg{} $=3.87^{+0.07}_{-0.08}$~dex, and $\text{R}_\text{p}=1.50\pm{+0.03}$~\rjup. While all three models are shown in Figure~\ref{fig:seds_main}, we lean on the BT-SETTL atmospheric model for the majority of this work to maintain consistency with \cite{carter2023}. We observe characteristic near-IR absorption features of water and CO along with a thermal continuum typical of a young, cloudy, giant planet atmosphere. Compared to field brown dwarfs of similar effective temperature, HIP~65426~b differs by exhibiting a triangular H-band peak around 1.65~\um~and comparatively weaker CH$_4$ absorption near 1.6~\um~and 2.2~\um, consistent with young, low-surface-gravity objects \citep{allers2013}.

Our astrometry and photometry values are shown in Table~\ref{tab:table_b}. They are consistent with the values in \cite{carter2023} to within the sigma values indicated in Table~\ref{tab:sigma_dif}. Starting with separation, NIRCam ranges from $0.0-1.3\sigma$ and is in good agreement. For MIRI, it is $0.25\sigma$ for F1140C and $2.21\sigma$ for F1550C. Updates to the calibration files could play a role in this discrepancy as well as small changes in image registration, reference star alignment, and KLIP parameters. These effects are more pronounced with broader PSFs which could account for this difference in F1550C. The position angles are consistent from $0.17-1.47\sigma$, with the largest difference in F250M. Since the F250M images were blurred, this could account for a less precise centroid. The most significant differences lie in the calculated flux. The flux values were closer to those in the original analysis if we employed a Matérn 3/2 kernel, but we chose to use the diagonal kernel for HIP~65426~b for consistency. In the case of MIRI, early photometric calibration files were underestimated by a factor of two and are a significant driver behind the differences in the MIRI fluxes\footnote{see MR-CI01, \url{https://jwst-docs.stsci.edu/known-issues/miri-known-issues/miri-coronagraphy-known-issues\#gsc.tab=0}}, as this issue was not resolved until after the initial analysis had been published. While we do not seek to diagnose the exact cause of the discrepancies due to many changes in all of the pipelines, software, and calibration files in the last three years, this work shows that these updates have indeed affected the photometric and astrometric calculations, specifically separation in F444W and F1550C, position angle in F250M, and flux in F250M, F300M, F410M, F1140C, and F1550C.

\begin{table}
    \centering
    \caption{Z-score for astrometric and photometric consistency with \cite{carter2023}. Values have a + or - sign to signify a positive or negative change.}
    \label{tab:sigma_dif}
    \begin{tabular}{llll}
\toprule
Filter & $\rho$ & $\theta$ & Flux \\
\midrule
F250M & -0.51 & +1.47 & +1.98 \\
F300M &  -0.82 & -0.33 & +3.69\\
F356W &  0.00 & -0.84 & +0.69\\
F410M &  -0.47 & -0.50 & +1.22\\
F444W &  -1.30 & +0.17 & +1.11\\
F1140C &  -0.25 & -0.78 & +1.91\\
F1550C &  -2.21 & +1.08 & +1.92\\
\bottomrule
\end{tabular}
   
\end{table}

\subsubsection{Background Stars}
\label{subsubsec: background stars}
Apart from HIP~65426~b, the eight point source CCs that fell within the TRILEGAL stellar population were well-matched to stellar atmosphere models, such as that shown for c1 in Figure~\ref{fig:seds_main}. They all displayed a smooth decrease in flux in the near-IR while flattening out toward the mid-IR that is characteristic of main-sequence stars. There was no evidence of absorption features or mid-IR excess that would be indicative of planetary or accreting objects. The slope of the SEDs from 2.5-4.5~\um~ corresponds with the expected Rayleigh-Jeans tail. The best-fit temperatures ranged from $\sim~2800-3800$~K with surface gravities characteristic of field stars, which is consistent with what we would expect from K-M dwarfs \citep{pecaut_mamajek2013, pecaut_mamajek_bubar2012}.

To investigate whether any of these background sources shared a common proper motion with HIP~65426~b, we compared their on-sky positions to those in the VLT/SPHERE data. This was done using \texttt{backtracks}\footnote{v0.6.4, \url{https://backtracks.readthedocs.io/en/latest/installation.html}} \citep{backtracks_code} and is shown in Figure~\ref{fig:backtracks_c5} in the appendix. We concluded that c5 likely correlates to cc-3 in Table F.2 in \cite{chauvin2017}. Consequently, the JWST observations provide an initial epoch of astrometry for the remaining sources that can be used for proper motion analysis in any future observations of this system.

\subsubsection{Other Sources}
\label{subsubsec: other sources}
In addition to the background stars and confirmed planet, we identified a small subset of sources that deviated from the expected SED patterns of stars or planets, specifically companion candidates c9, c10, c11, c18, and c20. c18 was outside the field of view in the NIRCam images, so the SED consists only of the two MIRI photometric points, which is not constraining enough to yield any definitive results from a model fit. On the other hand, c20 was in the NIRCam field of view but was just barely detected in two of the filters. The source appears very bright in F1550C, gets dimmer in F1440C, and is barely visible in a few NIRCam filters. A similar source was observed in \cite{boccaletti2024} and was determined to be a background galaxy. 

Due to the shape of MIRI PSFs, the extracted PSF of c20 appears as though it could be a point source, but it appears very spatially extended in F356W and F444W and is only marginally detected in these filters; therefore, the NIRCam measurements are treated as non-detections. Despite this, the only model that somewhat adequately fit the data was a blackbody curve, as is shown in Figure~\ref{fig:seds_main}; however, the blackbody curve still fails to capture the bright 15~\um~ flux. c9, c10, and c11 also could not be well fit to any of the models in \texttt{species}. The models consistently failed to capture the near-infrared data points, especially those in the F250M, as shown in Figure~\ref{fig:seds_others}.  

To go about classifying these non-stellar, non-planetary sources, we used galaxy SED templates obtained from the SWIRE library \citep{swire}. For each source and model, a best-fit scale factor was calculated using a $\chi^2$ minimization method, which considered the contribution from each observed wavelength. For the filters with non-detections, we used a $3\sigma$ upper-limit. We found that c9, c10, and c11 were all well fit to the spiral galaxy SED templates, whereas c20 was well fit to the Seyfert, starburst, and ultra-luminous infrared galaxies. c9 was well-fit with both the BT-SETTL and galaxy models; however, the PSF fit was improved with the Powell minimization method, which indicates that its PSF is extended and likely belongs to a galaxy rather than a star. The SEDs for c9, c10, and c11 along with their fits are shown in Figure~\ref{fig:seds_others}. 

While we are confident that these sources are extragalactic in nature, further analysis would be required for confirmation; however, we can be certain that they are not planetary. The nature of c18 remains uncertain, as we are unable to obtain upper-limits in NIRCam due to it being outside of the field of view. Using its angular separation of 11.5" and the HIP 65426 system distance of $107.49\pm{0.40}$~pc \citep{gaia2021}, c18 would have a physical separation of $\sim1240$~AU. While this is an unusually large separation and unlikely to be a bound planet, there are exceptions, as planets have been observed at larger separations (e.g. COCONUTS~2~b \citep{coconuts}). More observations would be required for confirmation.

\subsubsection{Background Source Likelihood}
\label{subsubsec: bg likelihood}
\begin{table}
    \centering
    \caption{The probability that a background source with the same photometry as a given candidate will appear within that separation of the host star, calculated with \texttt{SIDERIS}. A higher percentage indicates a higher probability of being a background source.}
    \label{tab:sideris}
    \begin{tabular}{ll|ll}
\toprule
Source & Probability & Source & Probability \\
\midrule
c1 & $64.5101\%$ & c8 & $90.9669\%$\\
c2 & $32.7196\%$ & c9 & $98.9993\%$\\
c3 & $11.3753\%$ & c10 & $42.3094\%$\\
c4 & $14.5694\%$ & c11 & $54.2165\%$\\
c5 & $89.9052\%$ & c12 & $99.9806\%$\\
c6 & $21.7309\%$ & c20 & $>76.5887\%$\\

\bottomrule
\end{tabular}
   
\end{table}

We assessed the likelihood that each CC was a background object with a chance alignment, the results of which are shown in Table \ref{tab:sideris}. This was done using \texttt{SIDERIS}\footnote{\url{https://github.com/aiza-kenzhebekova/SIDERIS}}, which determines the likelihood that a background object with the same photometry as the candidate will appear within the specified separation of the host star. This is done using a Poisson approach to calculate probability, comparing the photometry of each candidate in the F356W and F444W filters to that of the JADES galaxies and the simulated TRILEGAL population. The probability for c20 represents a lower limit, as the photometry values in those filters are upper limits, and the JADES galaxies were not observed in the relevant MIRI filters. c18 was not included, as we do not have NIRCam photometry for this source. The probabilities range from $11.38\%$ to $99.99\%$ with the majority falling above $50\%$. Those falling below $50\%$ are the brightest and bluest sources. Since \texttt{SIDERIS} obtains the expectation value by calculating the number of objects that are brighter and redder than a given candidate, the brightest objects tend to have lower probabilities; however, even a probability of $~10\%$ is indicative of a background source, as true planet candidates typically have a probability of $\ll1\%$, such as the $0.005\%$ reported in \cite{bowens_rubin2026}.

\section{Conclusions} \label{sec:conclusion}
In this study, we analyzed 21 potential CC sources detected in the high contrast imaging data of HIP 65426 using both MIRI and NIRCam data. Through a systematic vetting process, we determined that 13 of those sources were background sources unassociated with the HIP 65426 system. The other eight sources were discarded as being sources of noise, artifacts, cosmic rays, or only being present in one filter. In addition to these 13 sources, HIP~65426~b was consistent with previous results as a young, giant planet with a cloudy atmosphere and an atmospheric model effective temperature estimate of $\sim$1436 K. While our SED fit and recovered effective temperature, \logg, and radius are consistent with previous analyses, our photometry and astrometry indicate that updates to pipelines and calibration files have affected these measurements, most noticeably separation in F444W and F1550C, position angle in F250M, and flux in F250M, F300M, F410M, F1140C, and F1550C. Some of these causes include updates to the JWST pipeline, \texttt{STPSF}, \texttt{spaceKLIP}, and photometric calibration files, but the exact cause of these changes is beyond the scope of this work. 

Eight of the background sources were found to be background stars with SEDs and effective temperatures characteristic of K-M dwarfs, and there is no evidence of them being associated with the HIP 65426 system. We conducted a search for common proper motion and determined that c5 is likely cc-3 from \cite{chauvin2017}. Therefore, this study provides an initial epoch of astrometry for the remaining CCs for future proper motion analyses. 

Additionally, we identified four likely extragalactic sources, particularly c20, although further investigation is required for confirmation. This source appears very visually extended in F356W and F444W but point-like in F1550C. This additional context indicates that we should be cautious of individual point source detections with MIRI imaging in a blind companions search since, as shown in this work, it may lead to a false positive determination of a point source object such as a bound planet; however, there are exceptions where companions can be confirmed without a second epoch. One such example includes when there is astrometric evidence or orbital predictions for a companion (e.g. 14 Her c in \cite{bardalez_gagliuffi2025}, where point source photometry agrees with the orbital prediction). This point may also apply to Gaia planets targeted by JWST in the future \citep{lammers2026}. Furthermore, it should be noted that this caution applies specifically to MIRI imaging, as the MIRI IFU could spectroscopically confirm a planet without a second epoch (e.g. Beta Pic d in \cite{gibbs2026}). While recent studies have demonstrated MIRI's increased sensitivities compared to NIRCam \citep{bowensrubin2025}, we must exercise caution in single-epoch detections due to its reduced ability to distinguish between point sources and extended objects. c20 shows the value in having multi-wavelength photometry from both NIRCam and MIRI, as it allows us to break the degeneracy in terms of PSF structure and color.  

When plotted on a color magnitude diagram, the eight stellar sources were consistent with the simulated TRILEGAL stars, and the likely galactic sources (apart from c18 and c20 due to lack of near-IR data) were consistent with the JADES galaxies, further supporting our classification of these objects. 

Overall, our findings give us a deeper understanding of the HIP 65426 system and its contents as well as provide a framework for background source characterization for JWST coronagraphic imaging data which can be applied to future JWST high contrast imaging programs.

\section*{Acknowledgments}
This work is based [in part] on observations made with the NASA/ESA/CSA James Webb Space Telescope. The data were obtained from the Mikulski Archive for Space Telescopes at the Space Telescope Science Institute, which is operated by the Association of Universities for Research in Astronomy, Inc., under NASA contract NAS 5-03127 for JWST. These observations are associated with ERS Program 1386.
The authors acknowledge the ERS Program 1386 team for developing their observing program with a zero-exclusive-access period. 
We thank the anonymous referee for their feedback, which improved the quality of this manuscript.
B.A.B. acknowledges funding by the UK Science and Technology Facilities Council (STFC) grant nos. ST/V000594/1 and UKRI1196.
S.H. acknowledges funding by the UK Science and Technology Facilities Council (STFC) grant nos. ST/Y002792/1.
B.J.S. acknowledges funding by the UK Science and Technology Facilities Council (STFC) grant nos. ST/V000594/1 and UKRI1196.
A.Z. acknowledges support from ANID -- Millennium Science Initiative Program -- Center Code NCN2024\_001 and Fondecyt Regular grant number 1250249.
C.L. acknowledges the financial contribution from PRIN MUR 2022 (code 2022YP5ACE) funded by the European Union – NextGenerationEU.

All the JWST data presented in this paper were obtained from the Mikulski Archive for Space Telescopes (MAST) at the Space Telescope Science Institute and can be accessed via DOI: \href{https://doi.org/10.17909/jkg1-4s10}{10.17909/jkg1-4s10}.

\clearpage
\bibliographystyle{aasjournal}
\bibliography{sample631}

\clearpage
\renewcommand{\thefigure}{A.\arabic{figure}}
\renewcommand{\thetable}{A.\arabic{table}}

\setcounter{figure}{0} 
\setcounter{table}{0} 
\appendix
\section{Companion Candidates} \label{sec:companion_candidates}
\renewcommand{\thefigure}{A.\arabic{figure}}
\renewcommand{\theHfigure}{A.\arabic{figure}}
\renewcommand{\thetable}{A.\arabic{table}}
\renewcommand{\theHtable}{A.\arabic{table}}
\begin{figure*}[h!]
    \centering
    \includegraphics[width=0.88\linewidth]{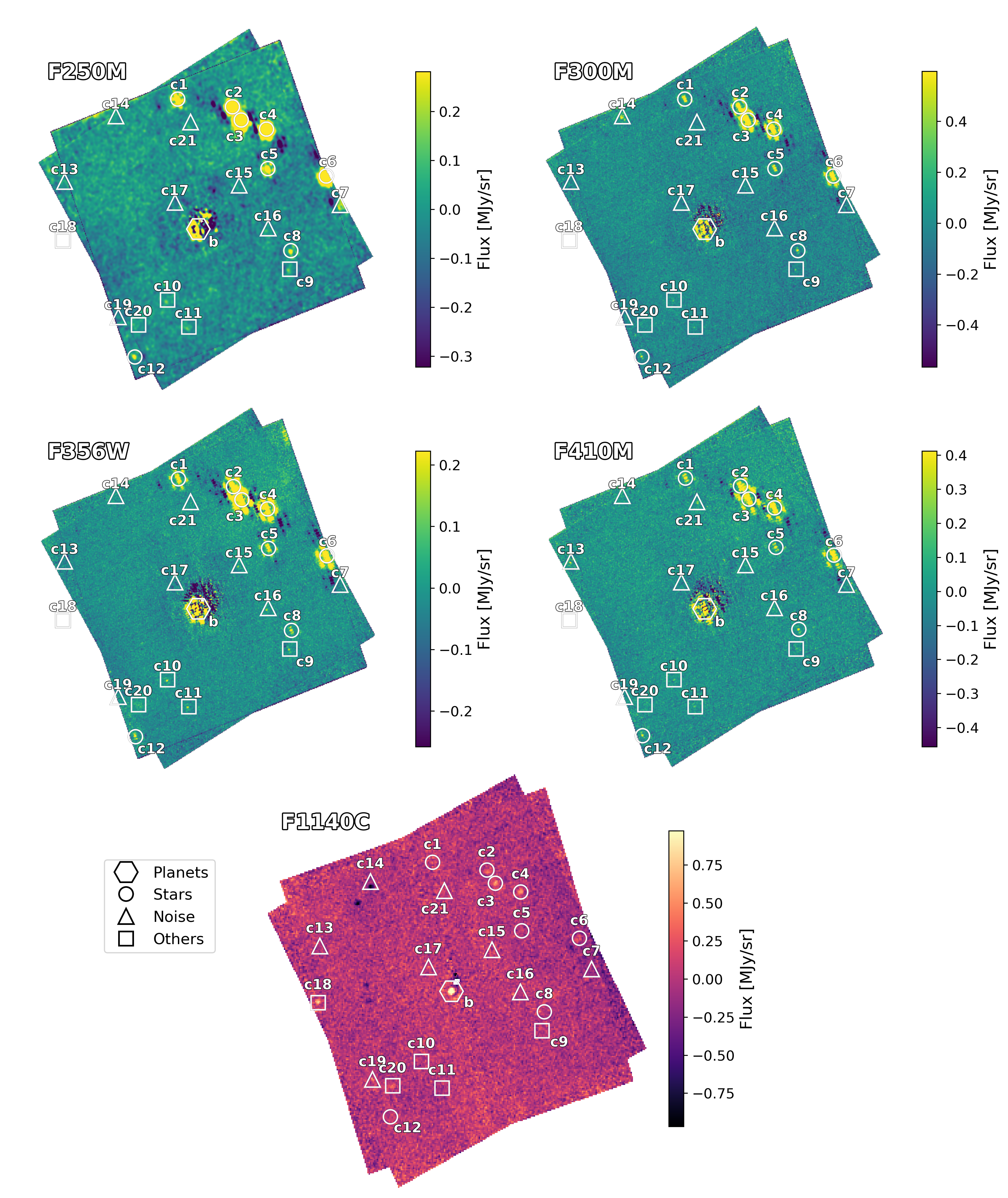}
    \caption{Remaining subtracted images not shown in Figure~\ref{fig:sources_f444_f1550}. All images are marked with their associated filter. The images are oriented with North pointing up and East pointing left. All candidate companion sources pre-vetting are marked with white markers. Not all sources were
    detected in all filters, so some markers appear empty.}
    \label{fig:sources_others}
\end{figure*}

 \clearpage 
\section{SED Fits}  \label{sec:sed_fits}
\renewcommand{\thefigure}{B.\arabic{figure}}
\renewcommand{\theHfigure}{B.\arabic{figure}}
\renewcommand{\thetable}{B.\arabic{table}}
\renewcommand{\theHtable}{B.\arabic{table}}

\setcounter{figure}{0} 
\setcounter{table}{0} 

\begin{figure}[h]
    \centering
    
    \begin{minipage}[t]{0.49\textwidth}
    \centering
    \includegraphics[width=\linewidth]{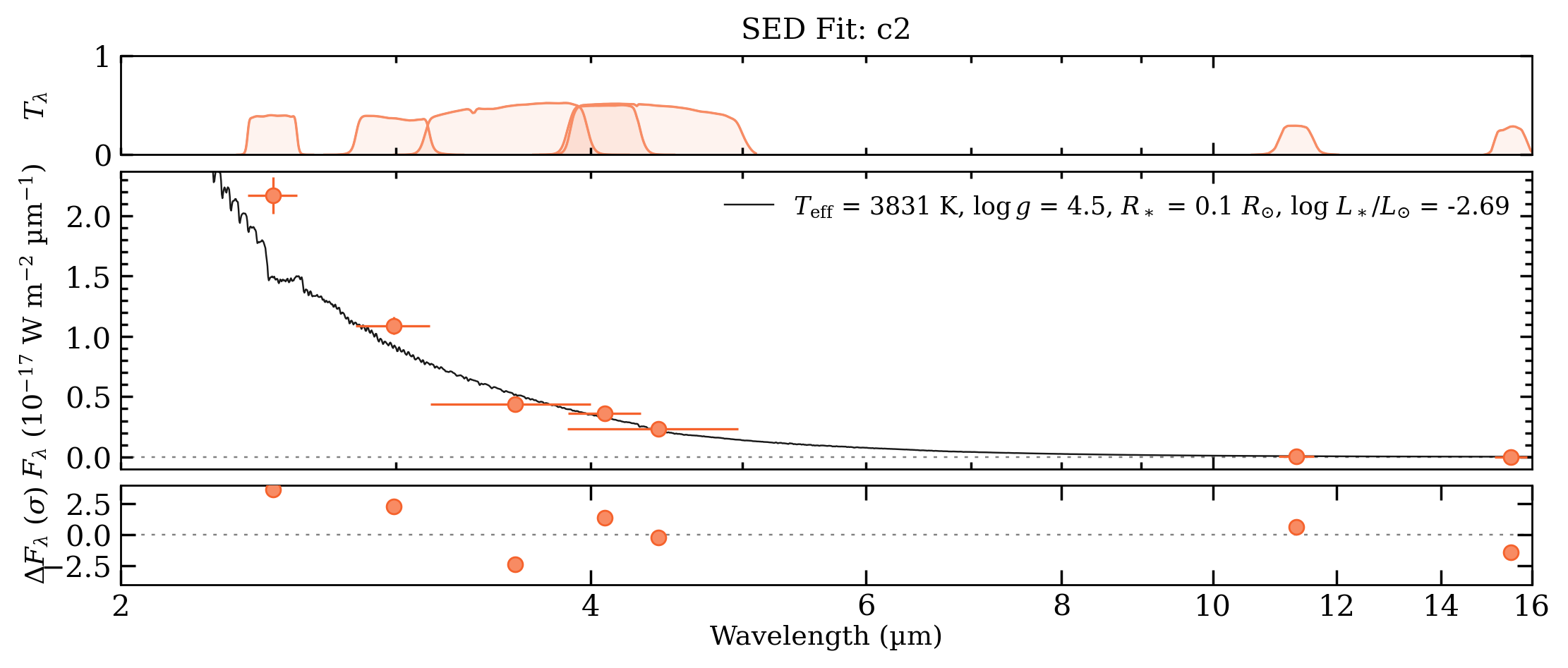}
    
    \includegraphics[width=\linewidth]{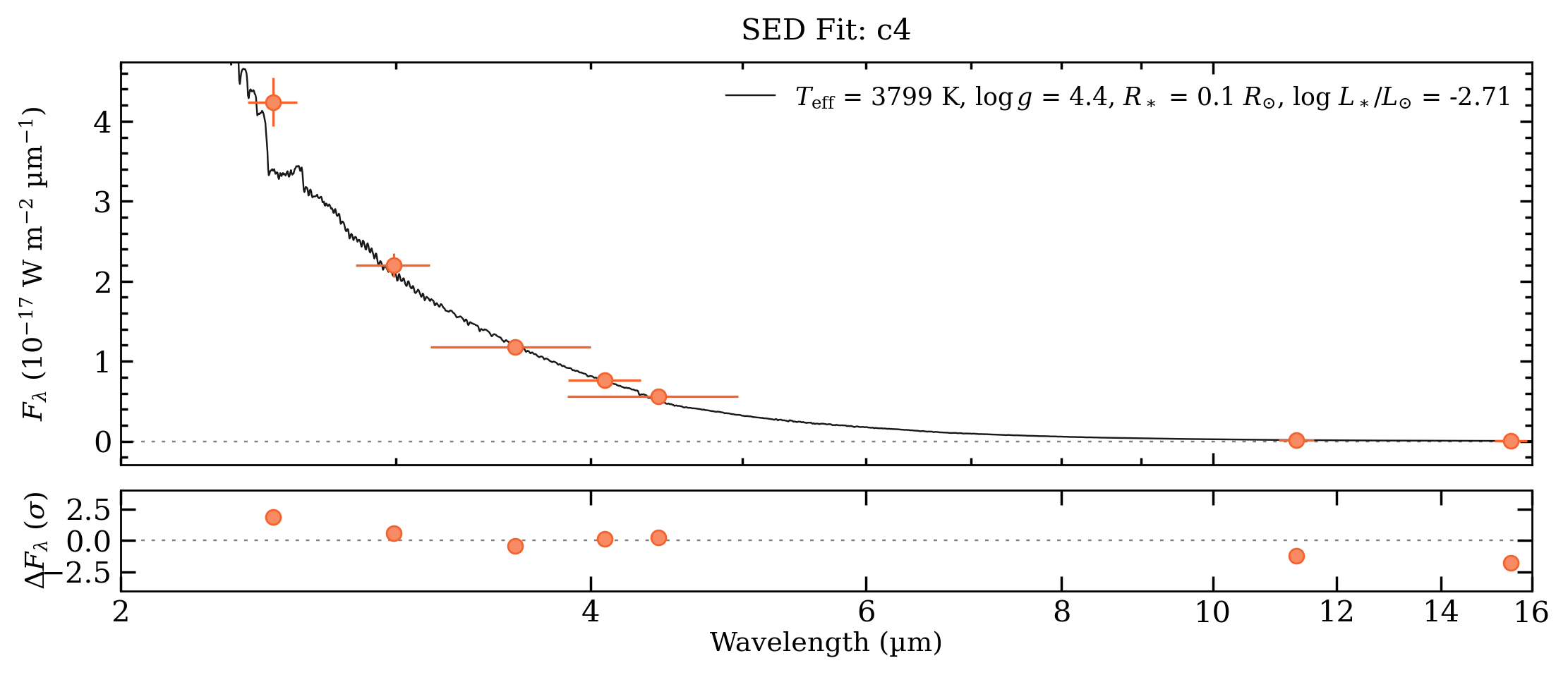}
    
    \includegraphics[width=\linewidth]{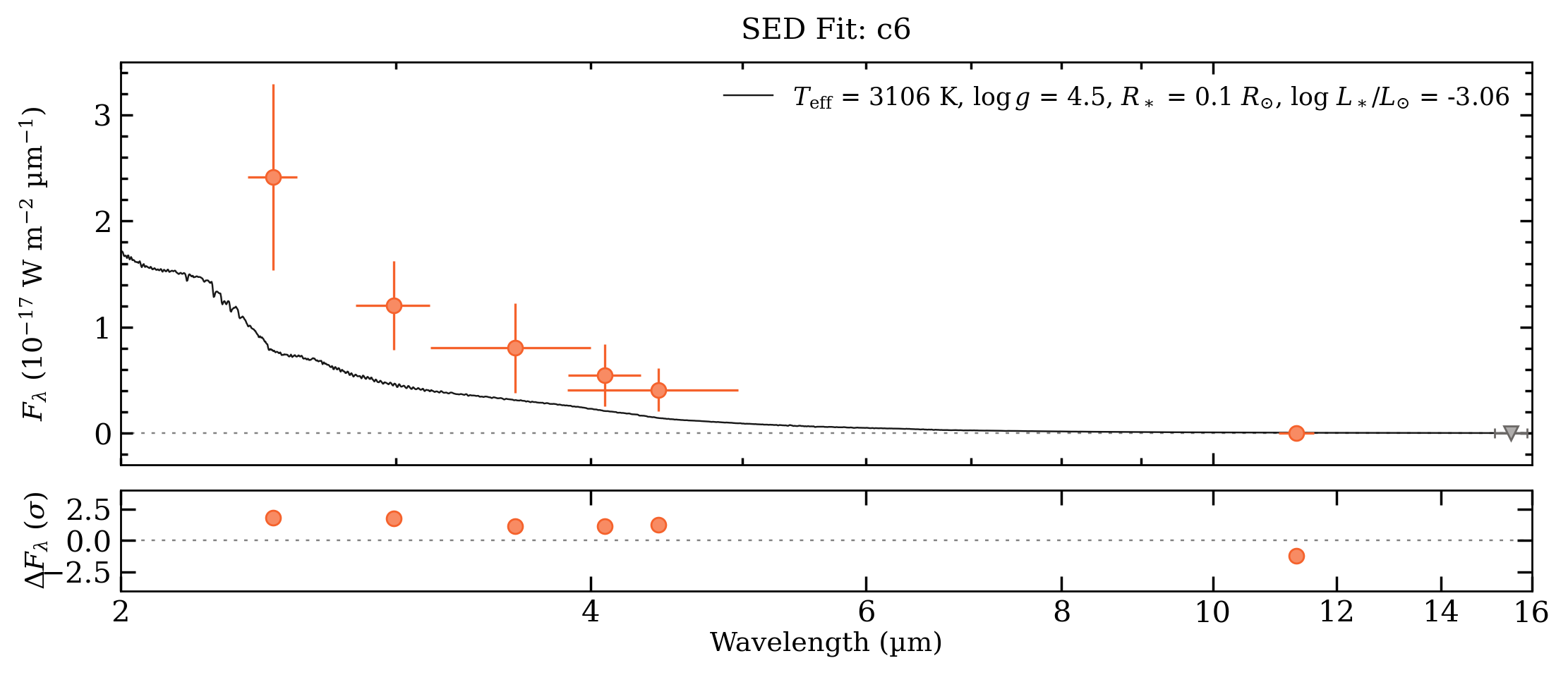}
    
    \includegraphics[width=\linewidth]{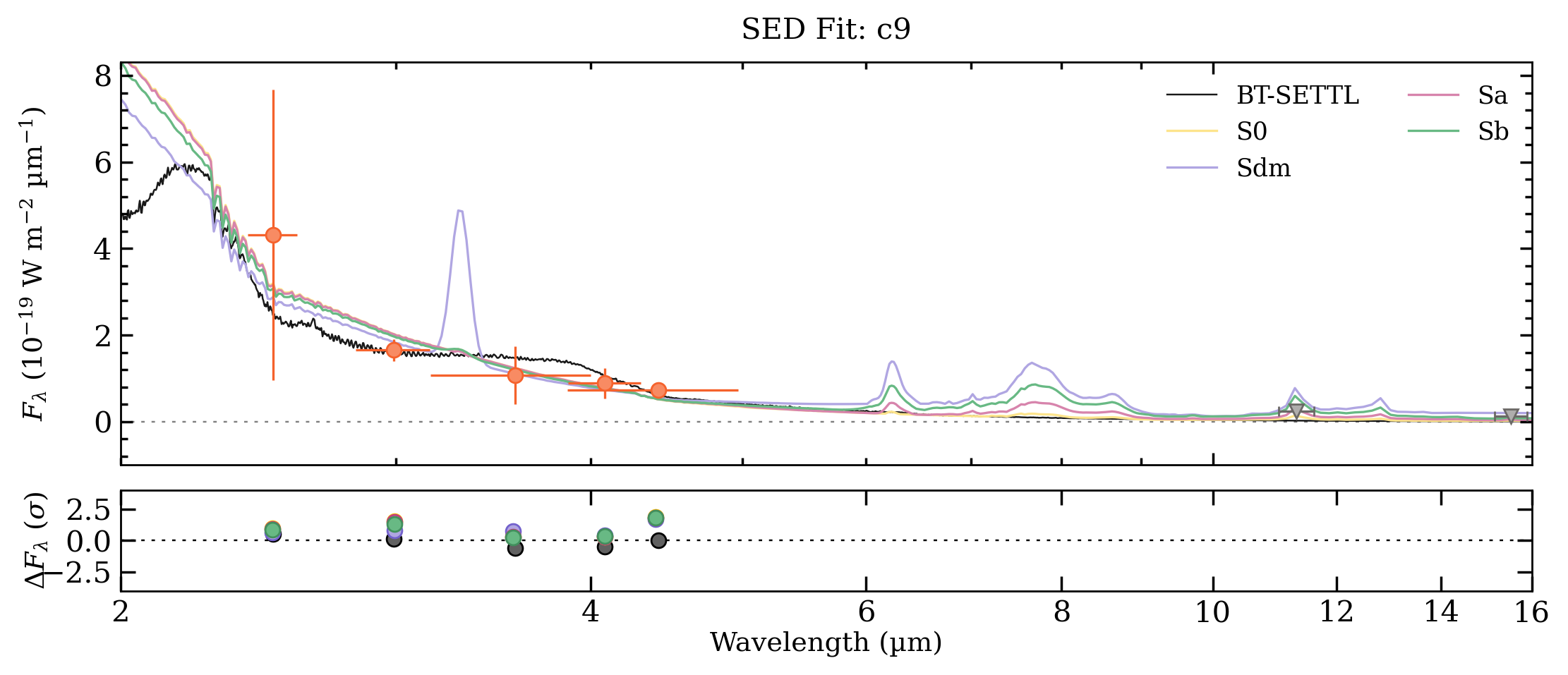}
    
    \includegraphics[width=\linewidth]{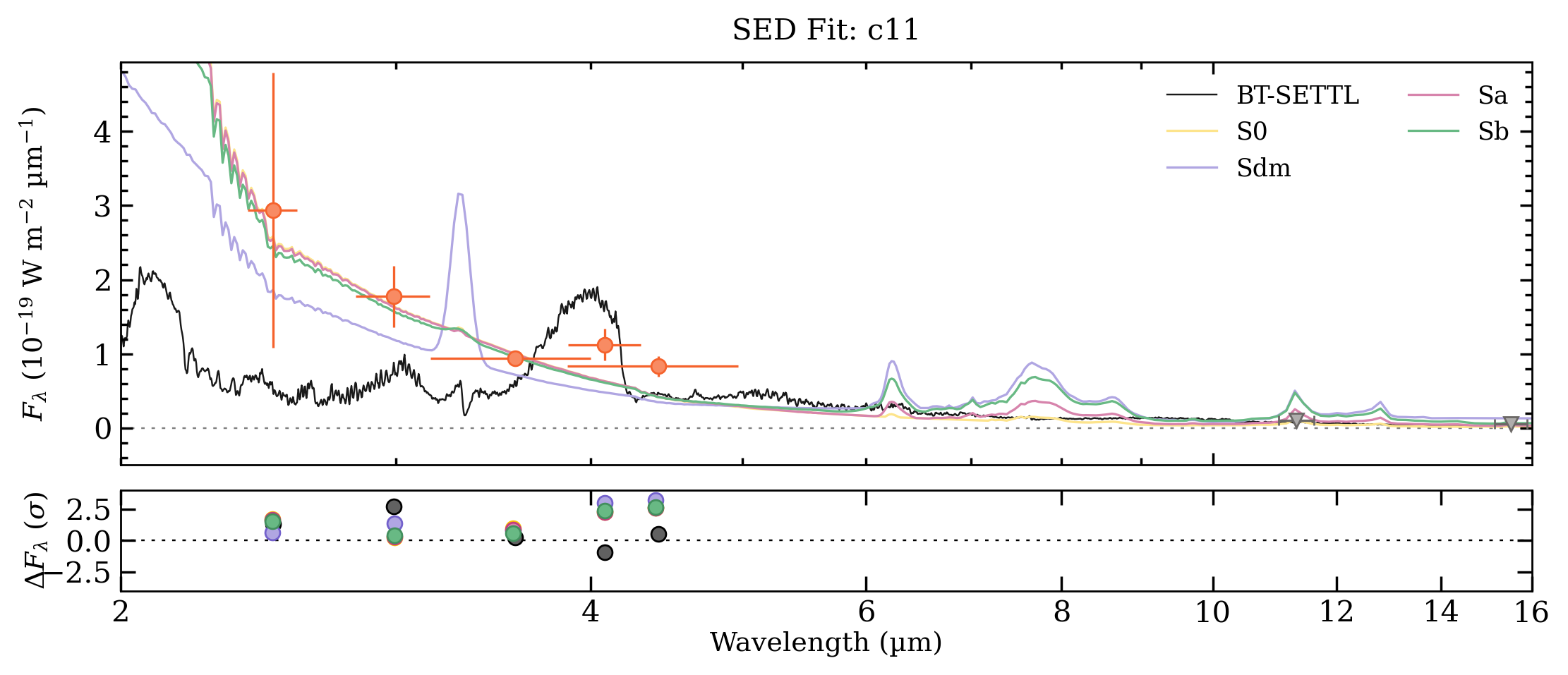}
    \end{minipage}
    \hfill
    \begin{minipage}[t]{0.49\textwidth}
    \centering
    \includegraphics[width=\linewidth]{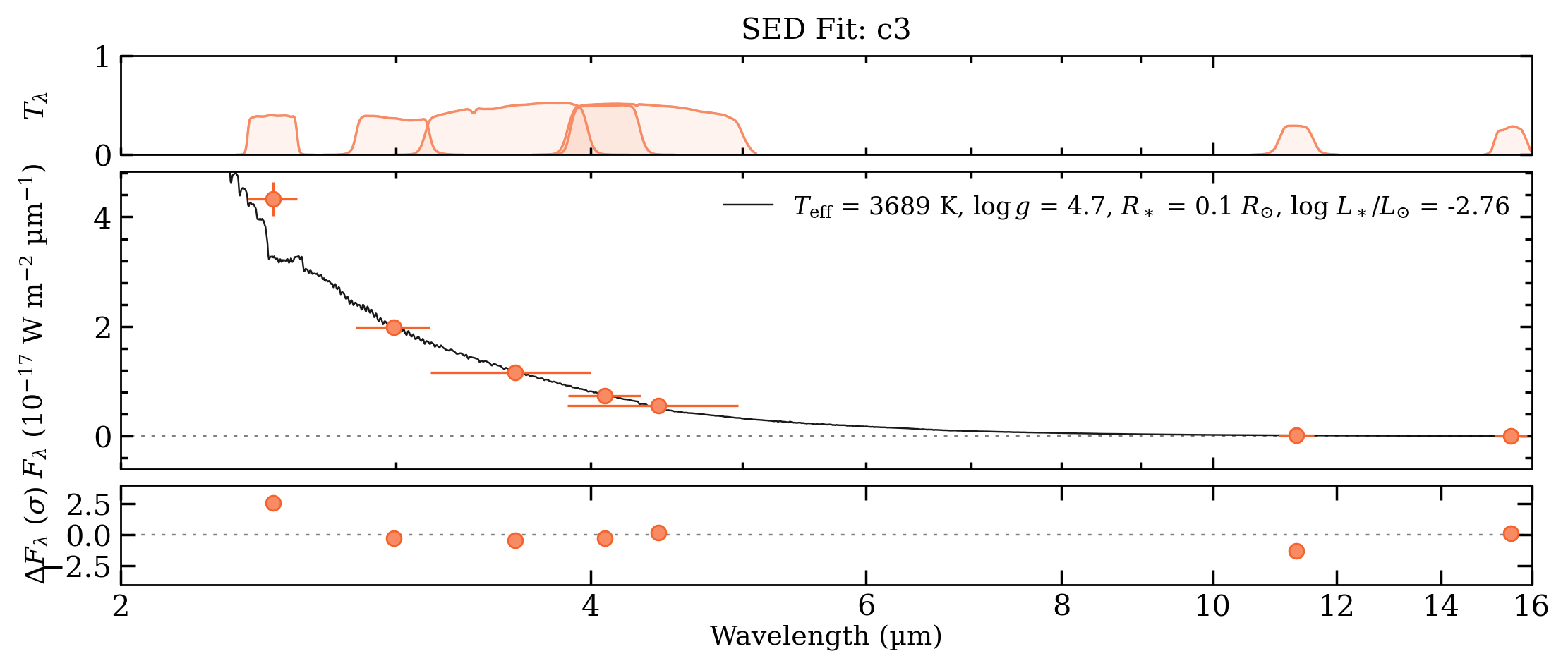}
    
    \includegraphics[width=\linewidth]{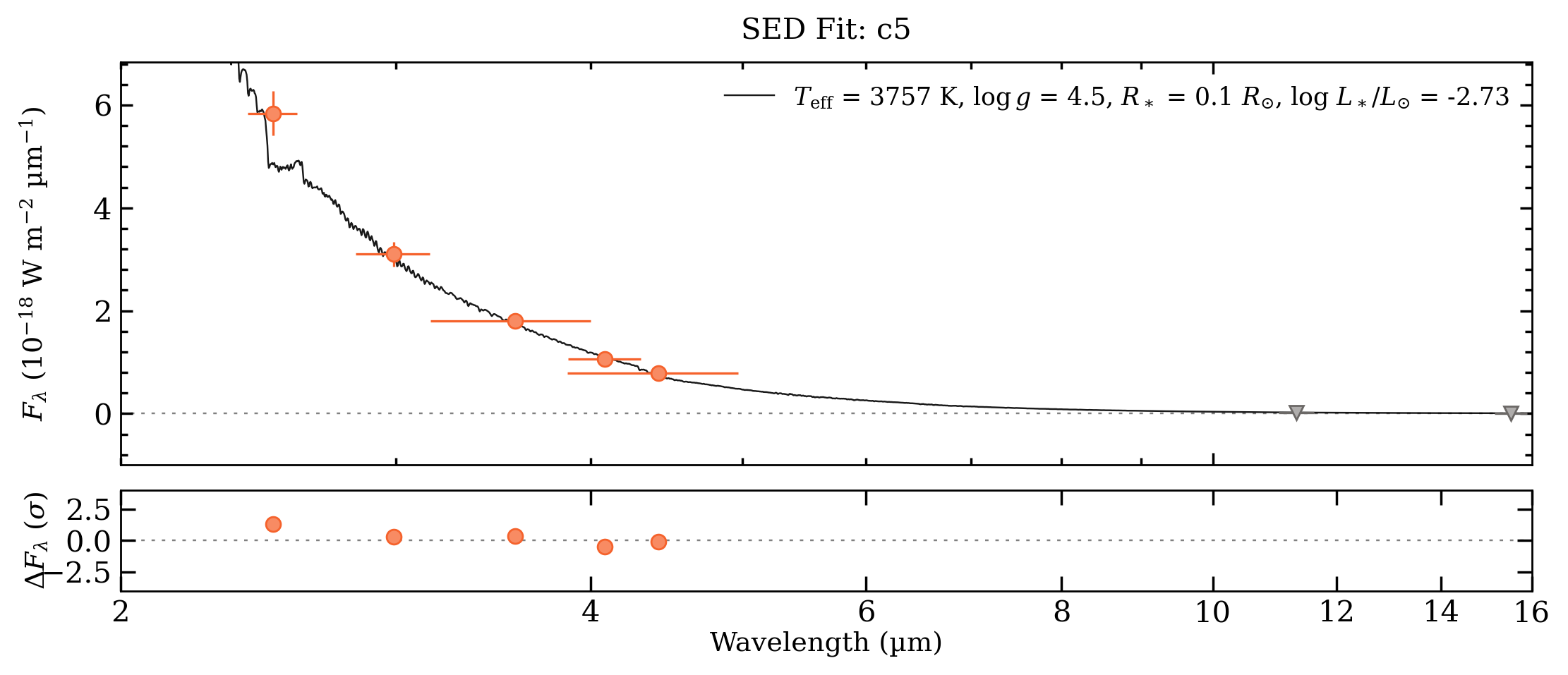}
    
    \includegraphics[width=\linewidth]{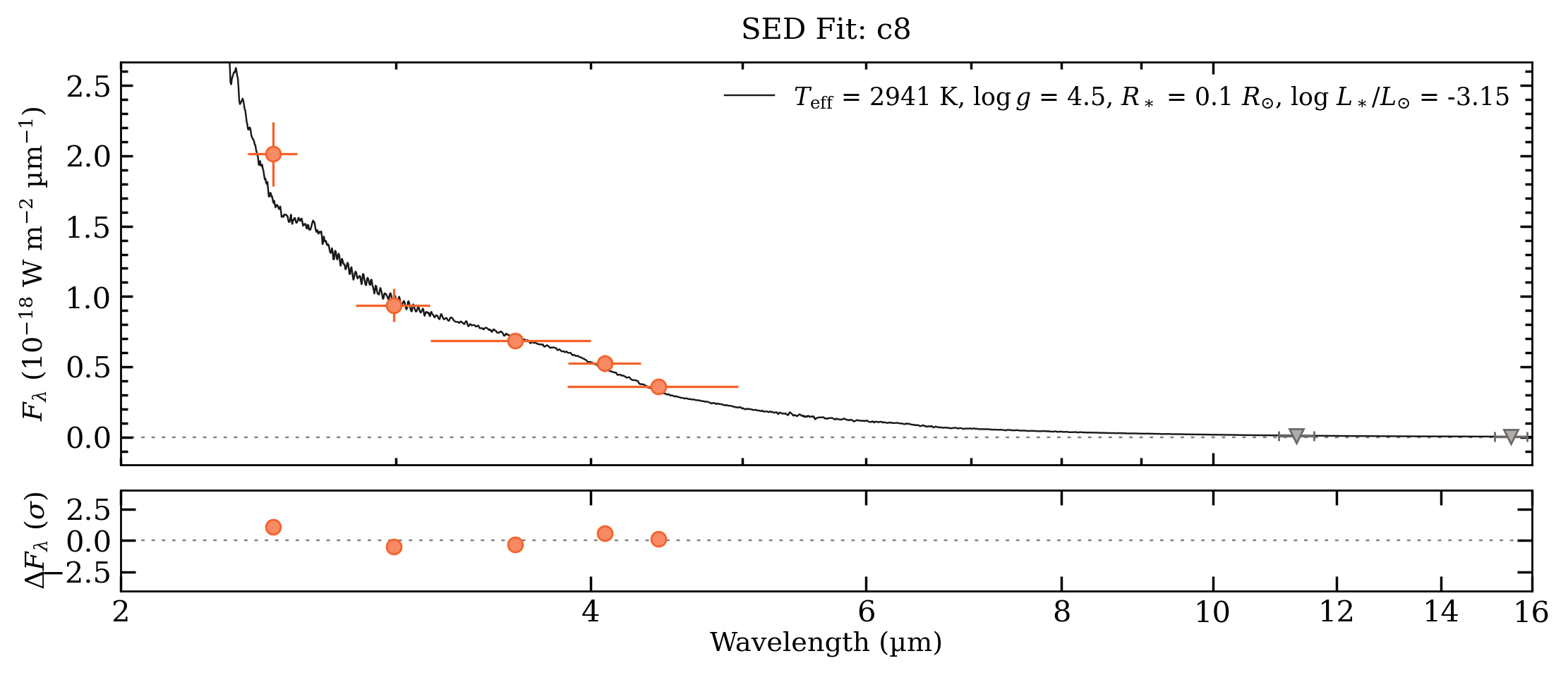}
    
    \includegraphics[width=\linewidth]{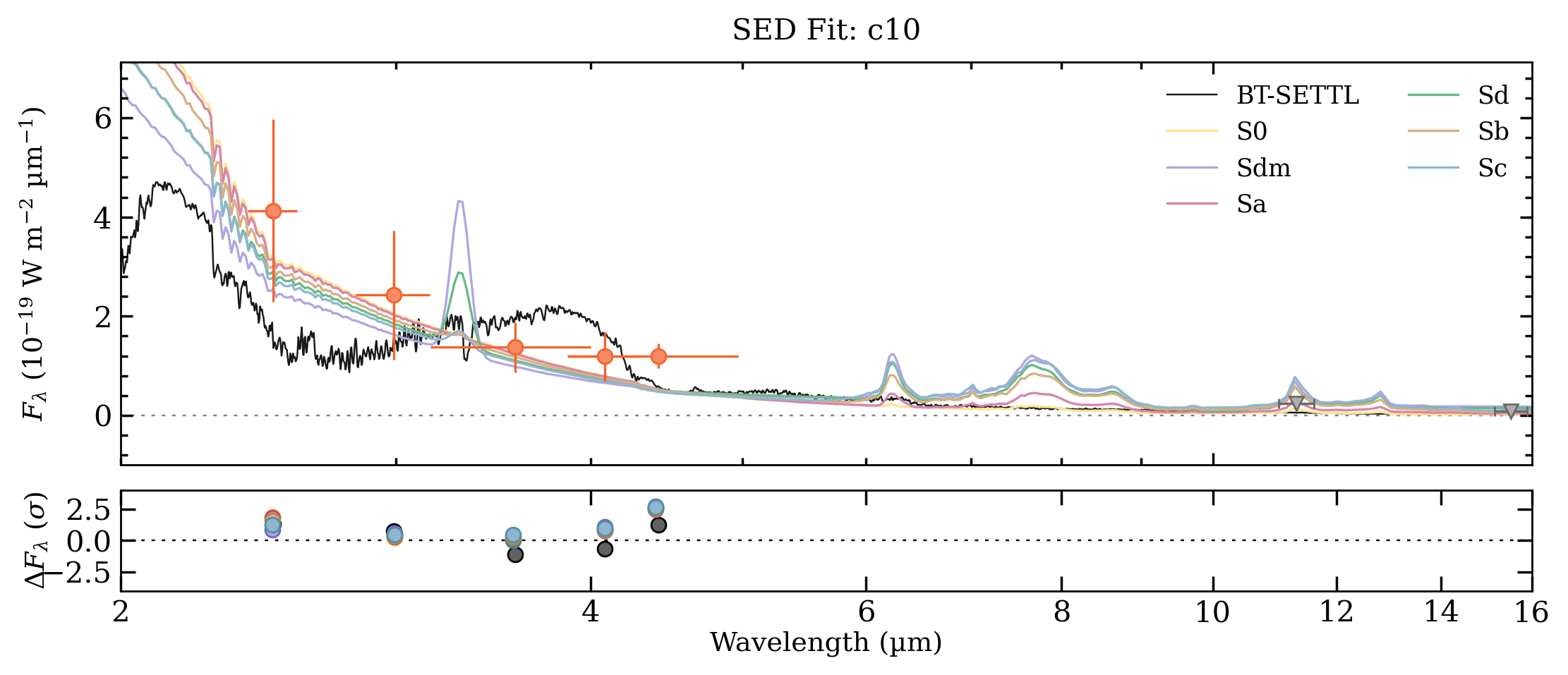}
    
    \includegraphics[width=\linewidth]{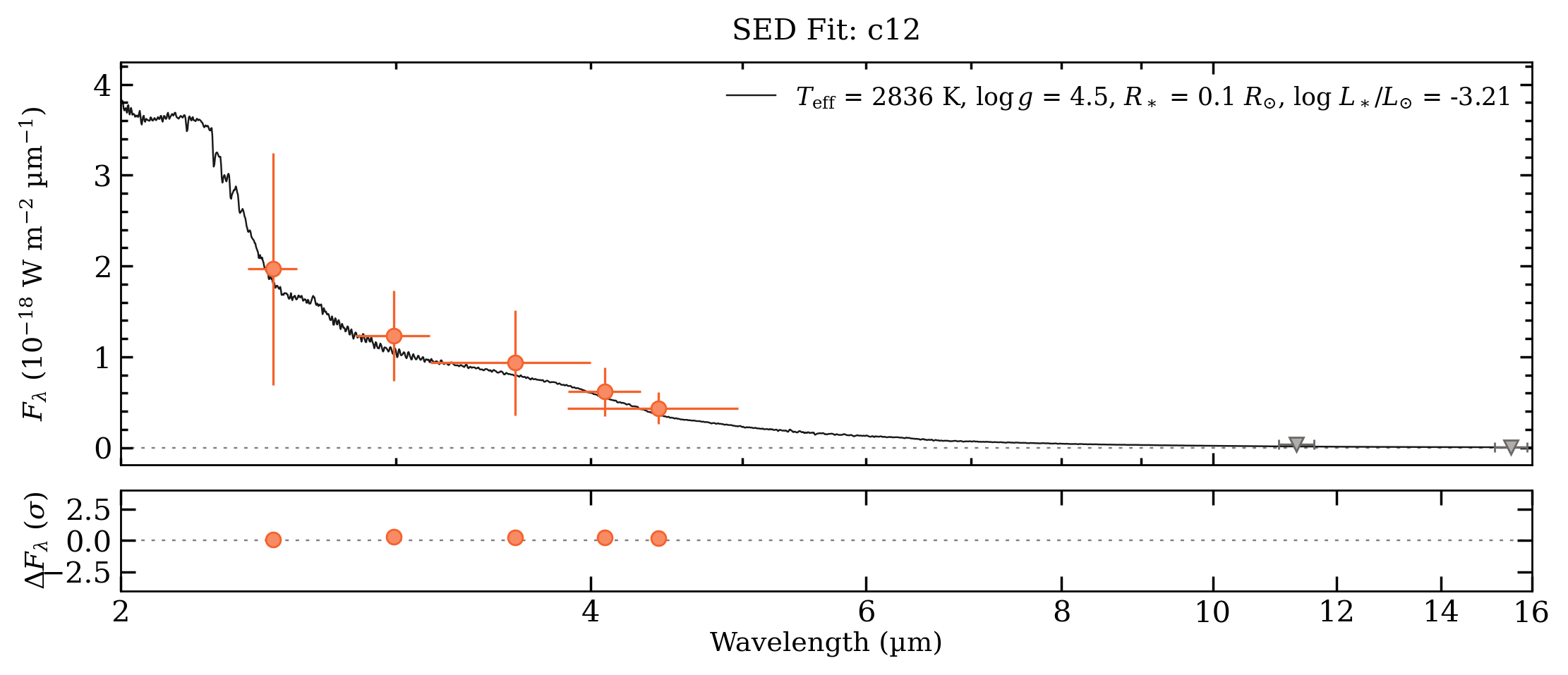}
    \end{minipage}

    \caption{SED fits for all vetted point-sources not included in Figure~\ref{fig:seds_main}. The JWST photometric points are plotted in orange with the best-fit BT-Settl models in black. c9, c10, and c11 also contain the best-fit galaxy SED templates. The residuals for each data point are shown in the bottom panel of each plot. 3$\sigma$ upper-limits are shown in grey.}
  \label{fig:seds_others}
\end{figure}

\clearpage
\section{Companion Fits}  \label{sec:companion_fits}
\renewcommand{\thefigure}{C.\arabic{figure}}
\renewcommand{\theHfigure}{C.\arabic{figure}}
\renewcommand{\thetable}{C.\arabic{table}}
\renewcommand{\theHtable}{C.\arabic{table}}
\setcounter{figure}{0} 
\setcounter{table}{0} 
The extracted fits for each companion in each filter are shown below. For each companion, the filters are in order with F250M at the top, then F300M, F356W, F410M, F444W, F1140C, and then F1550C at the bottom. 
\begin{figure}[h!]
    \centering
    \gridline{
    \fig{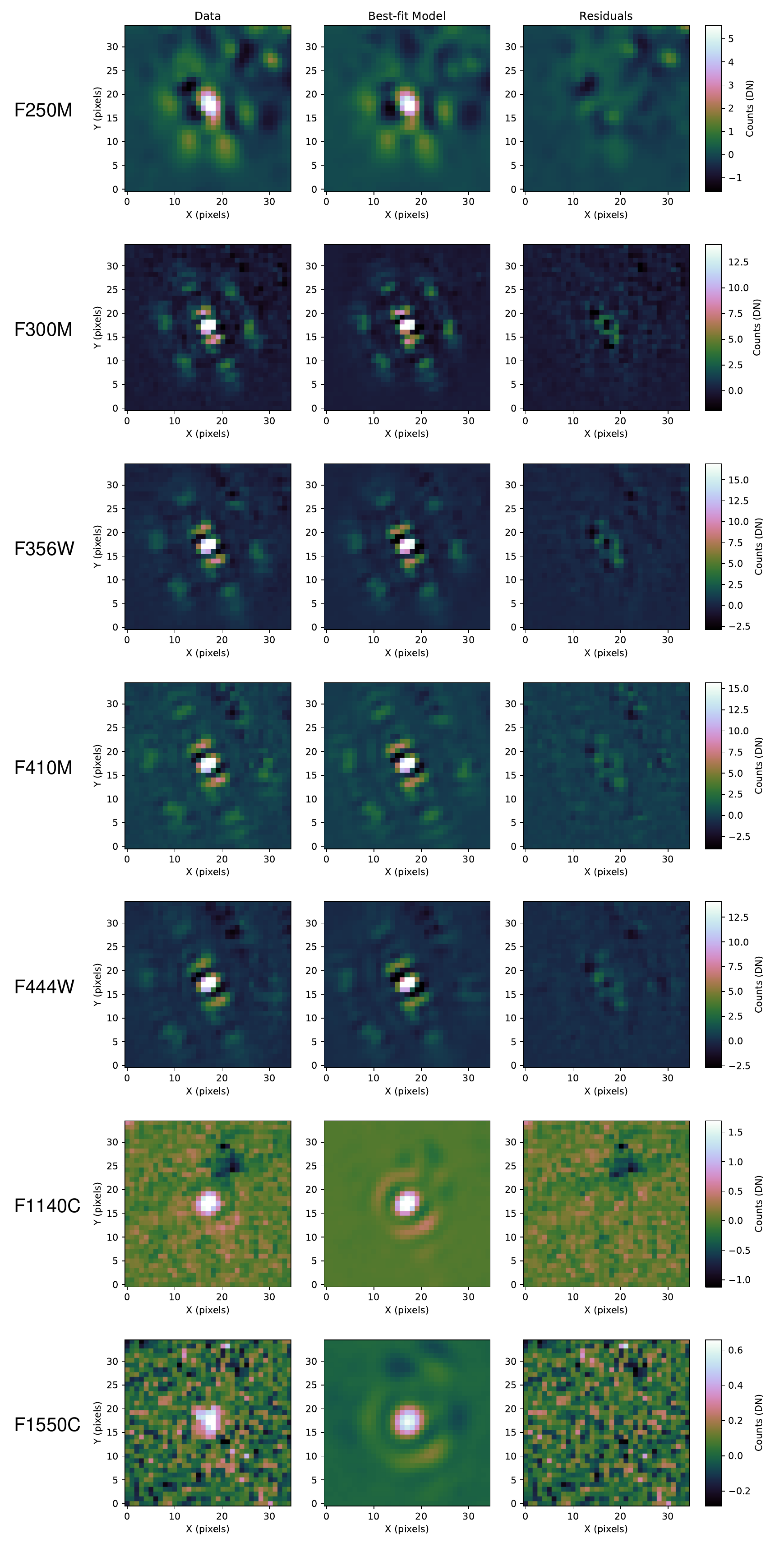}{0.5\linewidth}{(b)}
    \fig{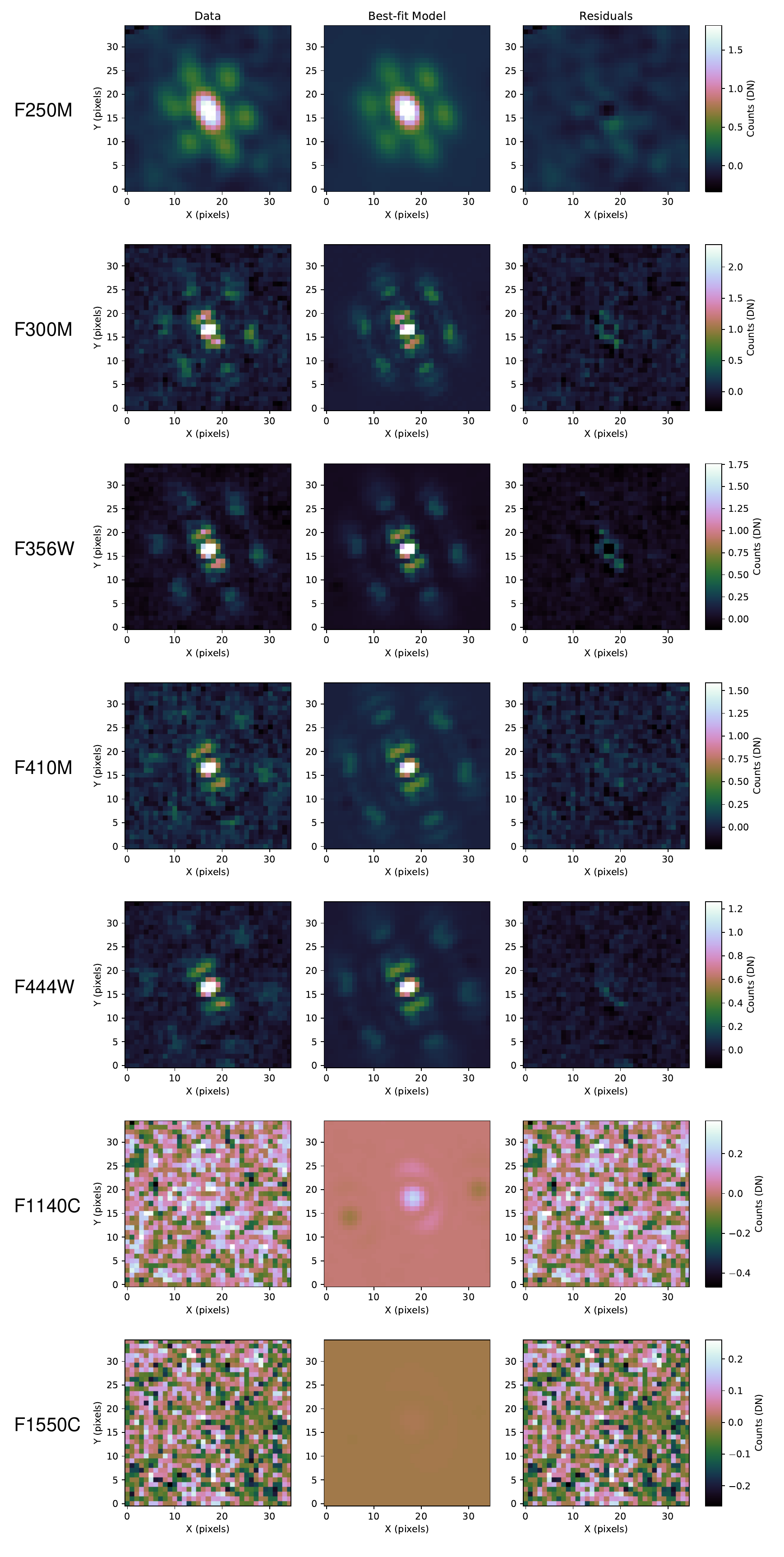}{0.5\linewidth}{(c1)}
    }
    \caption{Companion fits for b and c1.}
    \label{fig:psf_cbc1}
\end{figure}
\begin{figure}
    \figurenum{C.2}
    \centering
    \gridline{
    \fig{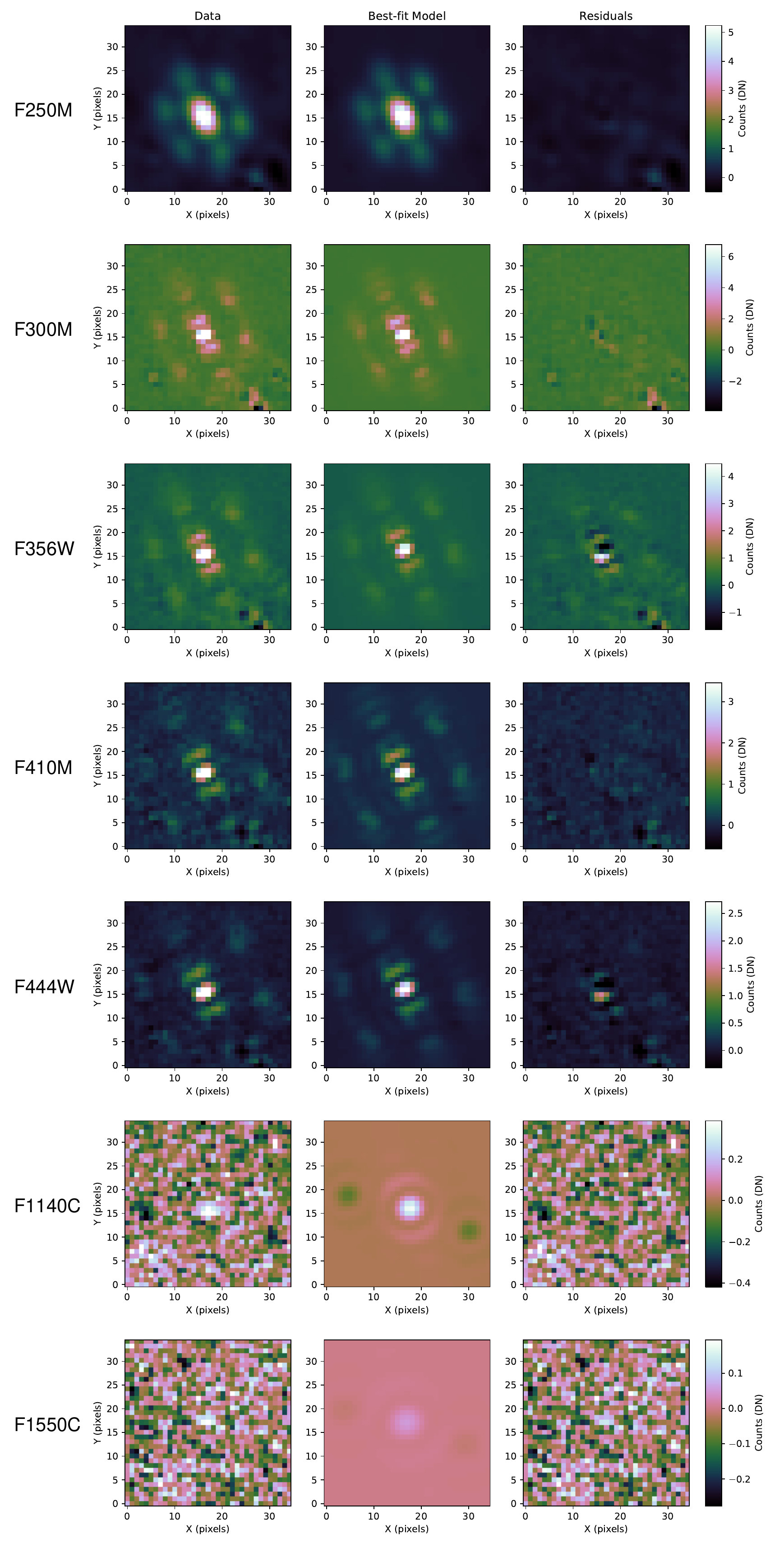}{0.5\linewidth}{(c2)}
    \fig{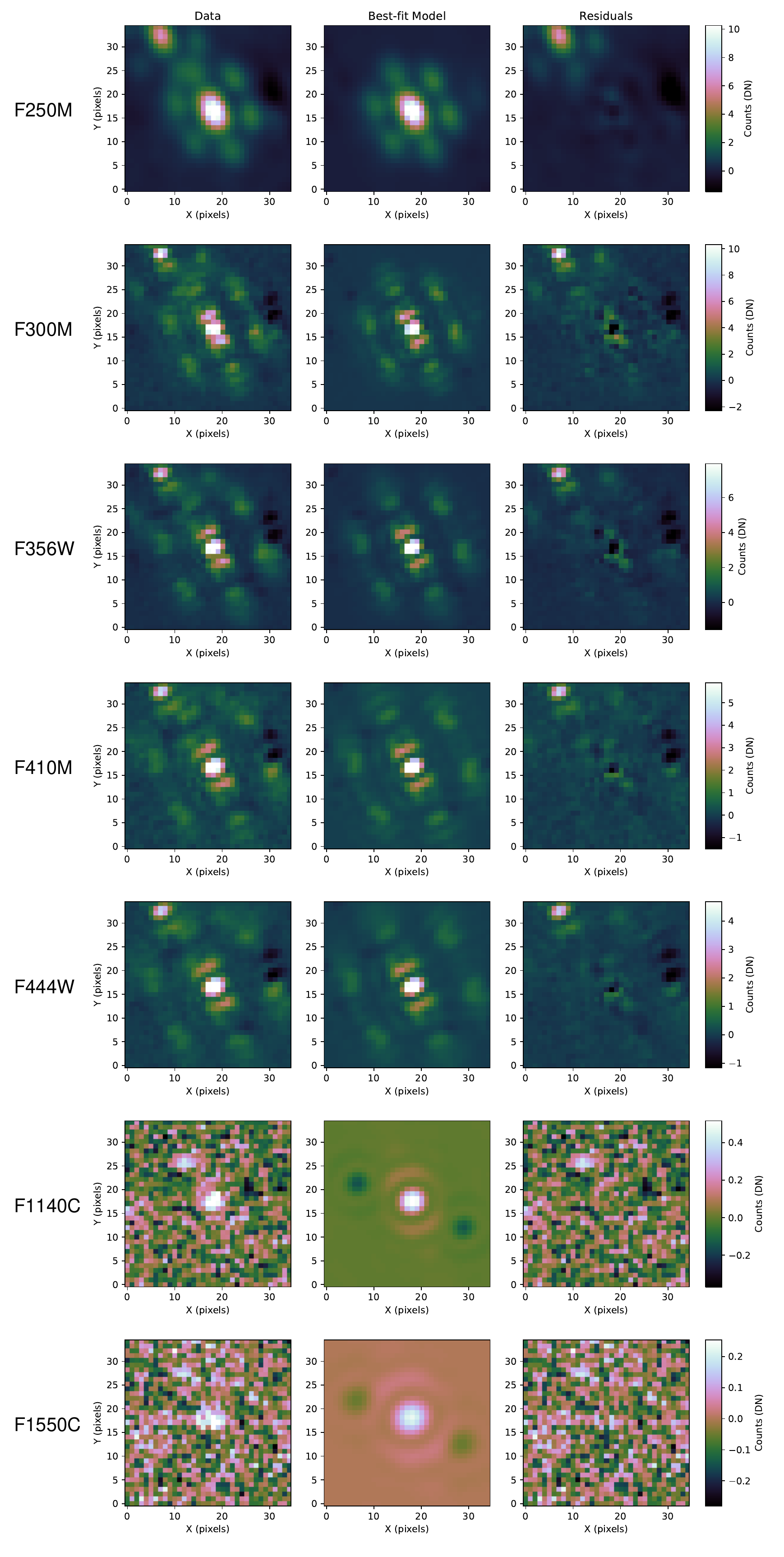}{0.5\linewidth}{(c3)}
    }
    \caption{Companion fits for c2 and c3.}
    \label{fig:psf_c2c3}
\end{figure}
\begin{figure}
    \figurenum{C.3}
    \centering
    \gridline{
    \fig{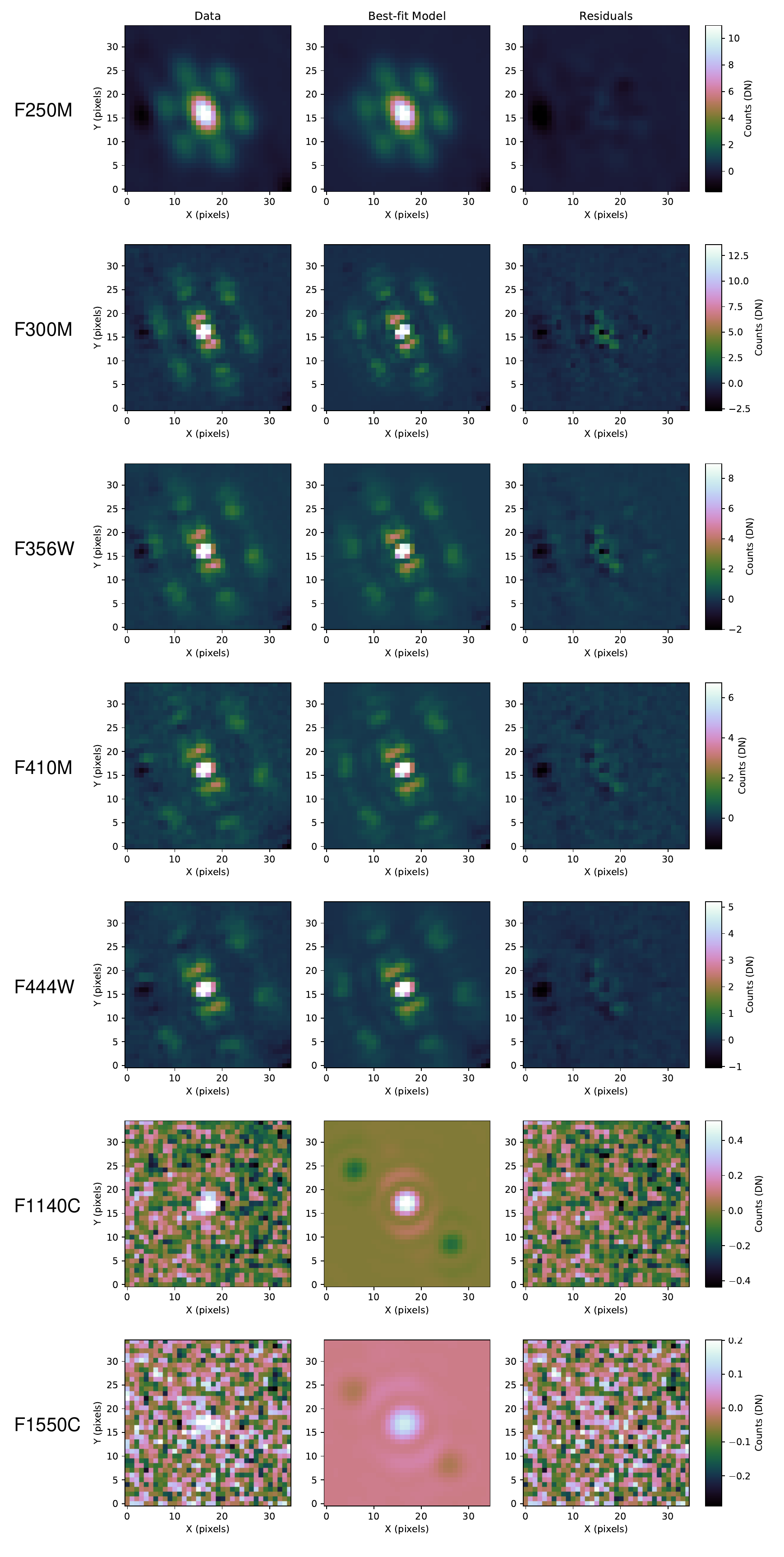}{0.5\linewidth}{(c4)}
    \fig{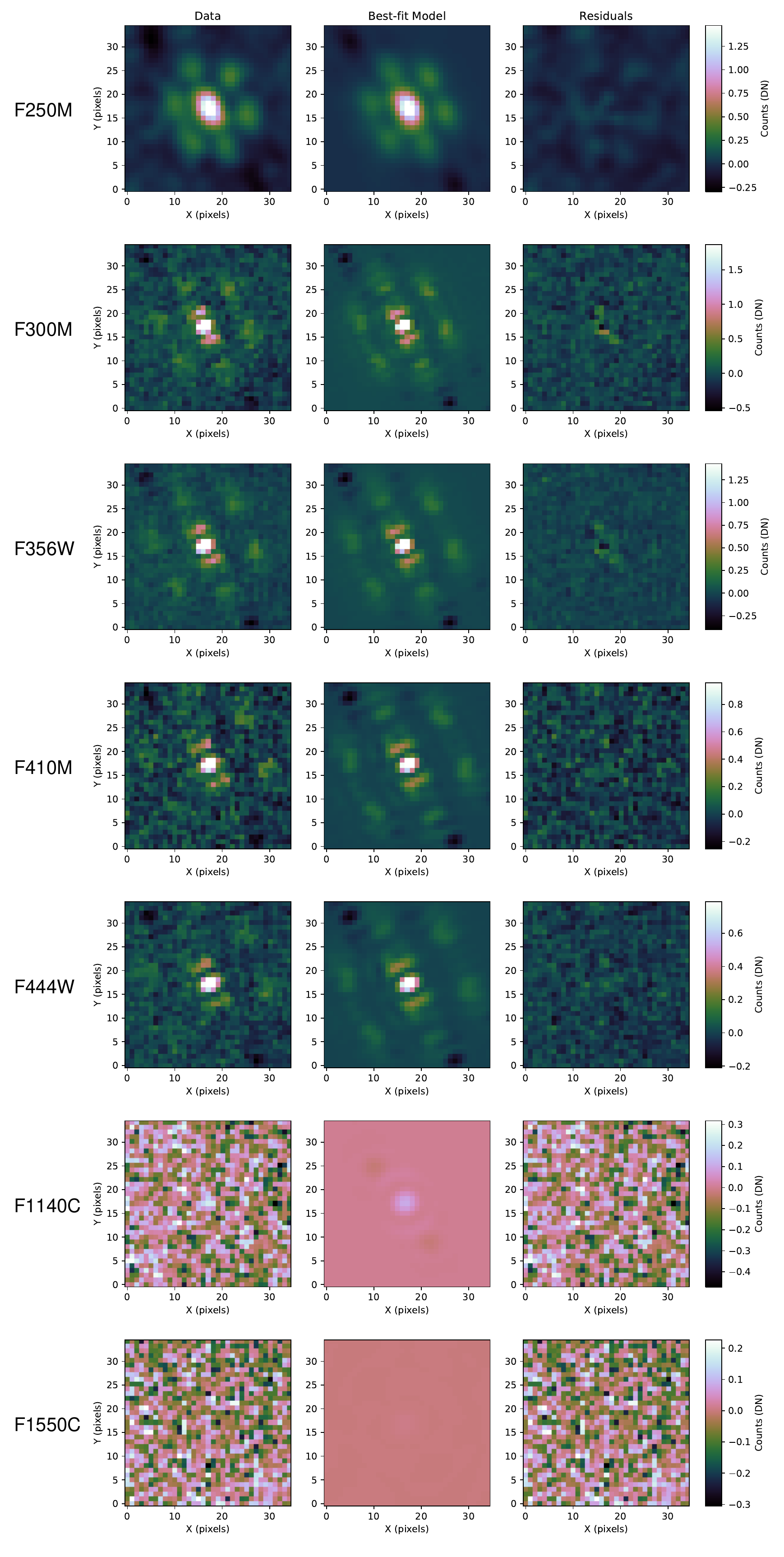}{0.5\linewidth}{(c5)}
    }
    \caption{Companion fits for c4 and c5.}
    \label{fig:psf_c4c5}
\end{figure}
\begin{figure}
    \figurenum{C.4}
    \centering
    \gridline{
    \fig{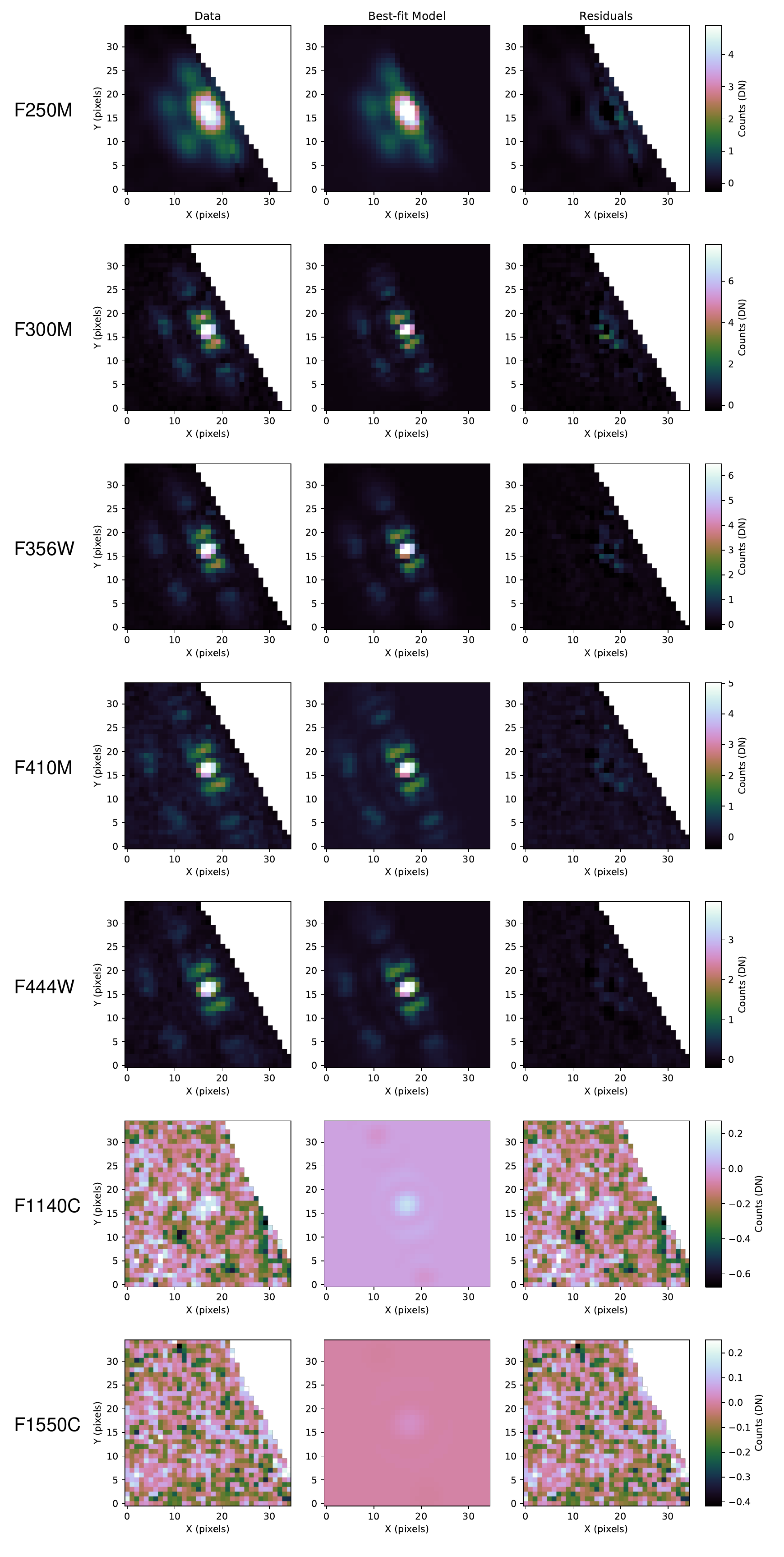}{0.5\linewidth}{(c6)}
    \fig{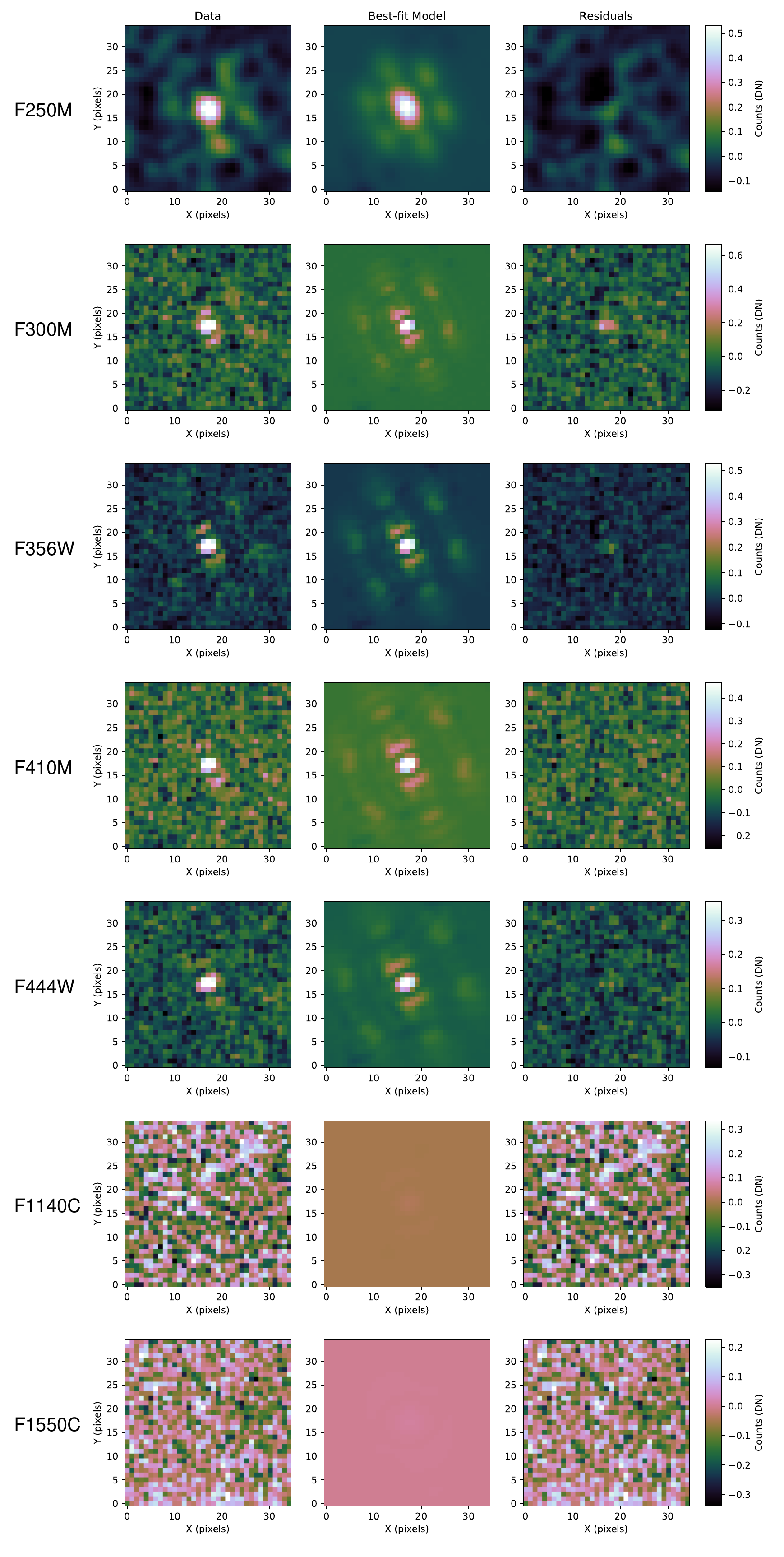}{0.5\linewidth}{(c8)}
    }
    \caption{Companion fits for c6 and c8.}
    \label{fig:psf_c6c8}
\end{figure}
\begin{figure}
    \figurenum{C.5}
    \centering
    \gridline{
    \fig{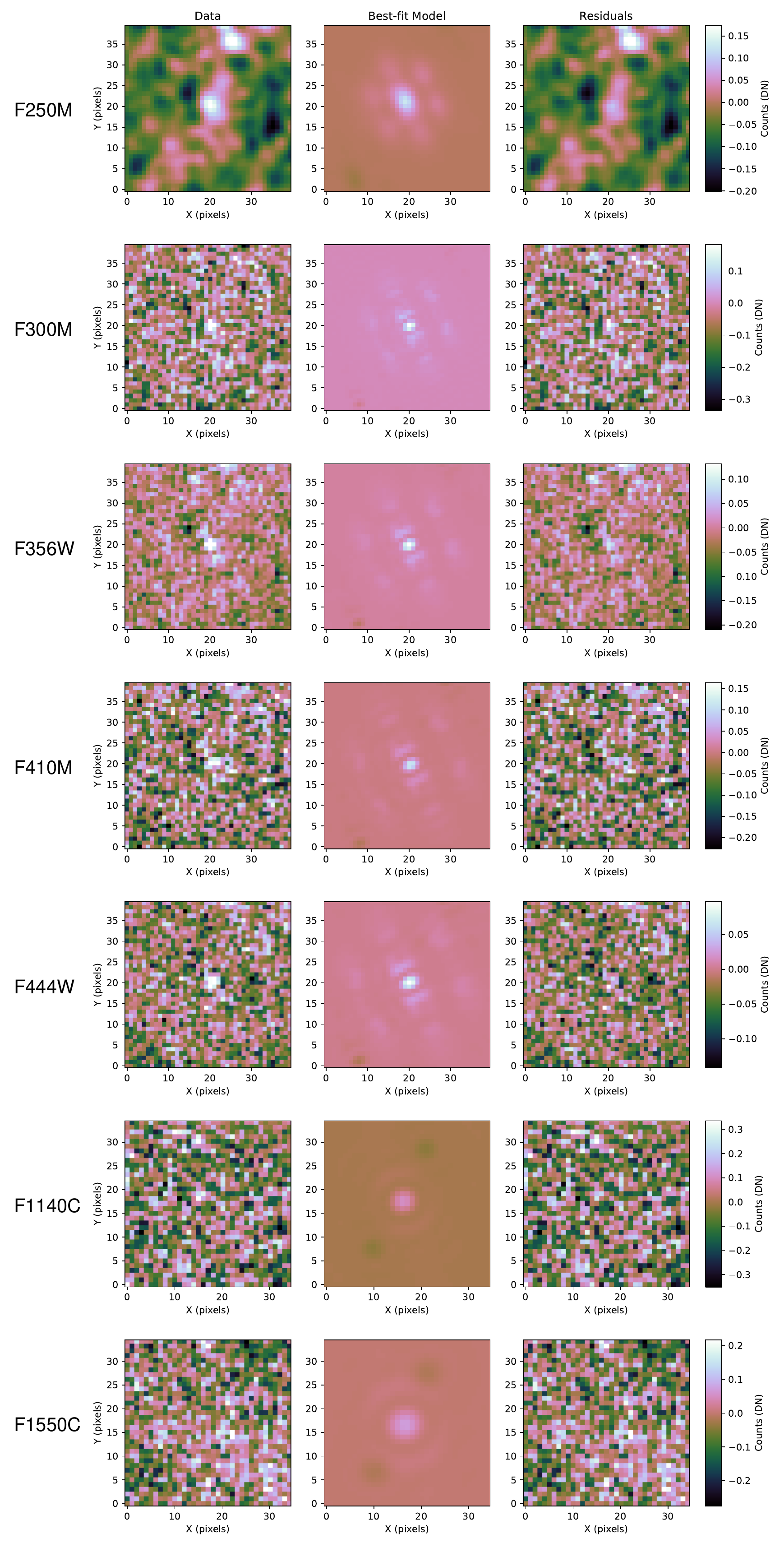}{0.5\linewidth}{(c9)}
    \fig{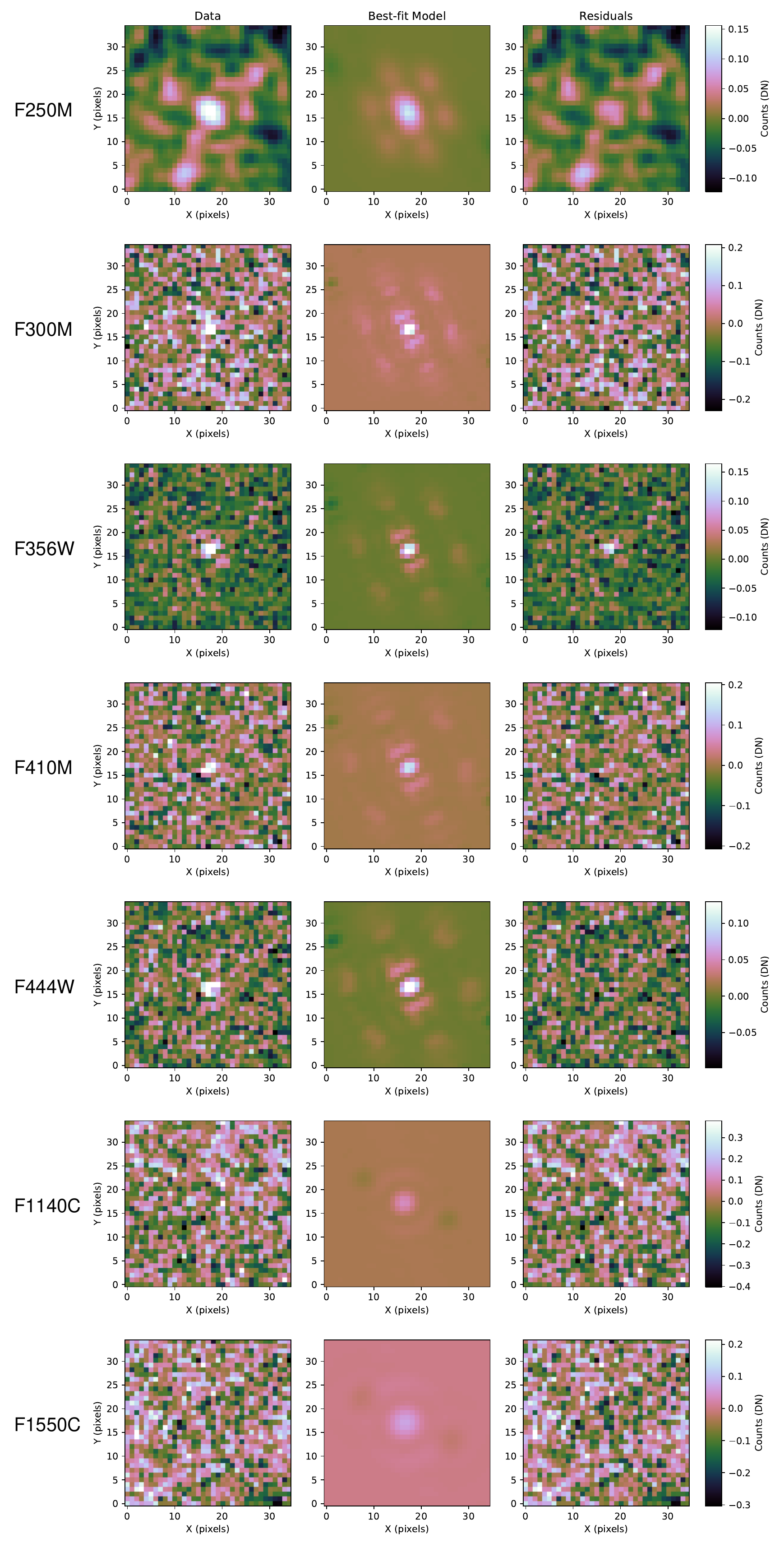}{0.5\linewidth}{(c10)}
    }
    \caption{Companion fits for c9 and c10 using the Powell minimization method}.
    \label{fig:psf_c9c10}
\end{figure}
\begin{figure}
    \figurenum{C.6}
    \centering
    \gridline{
    \fig{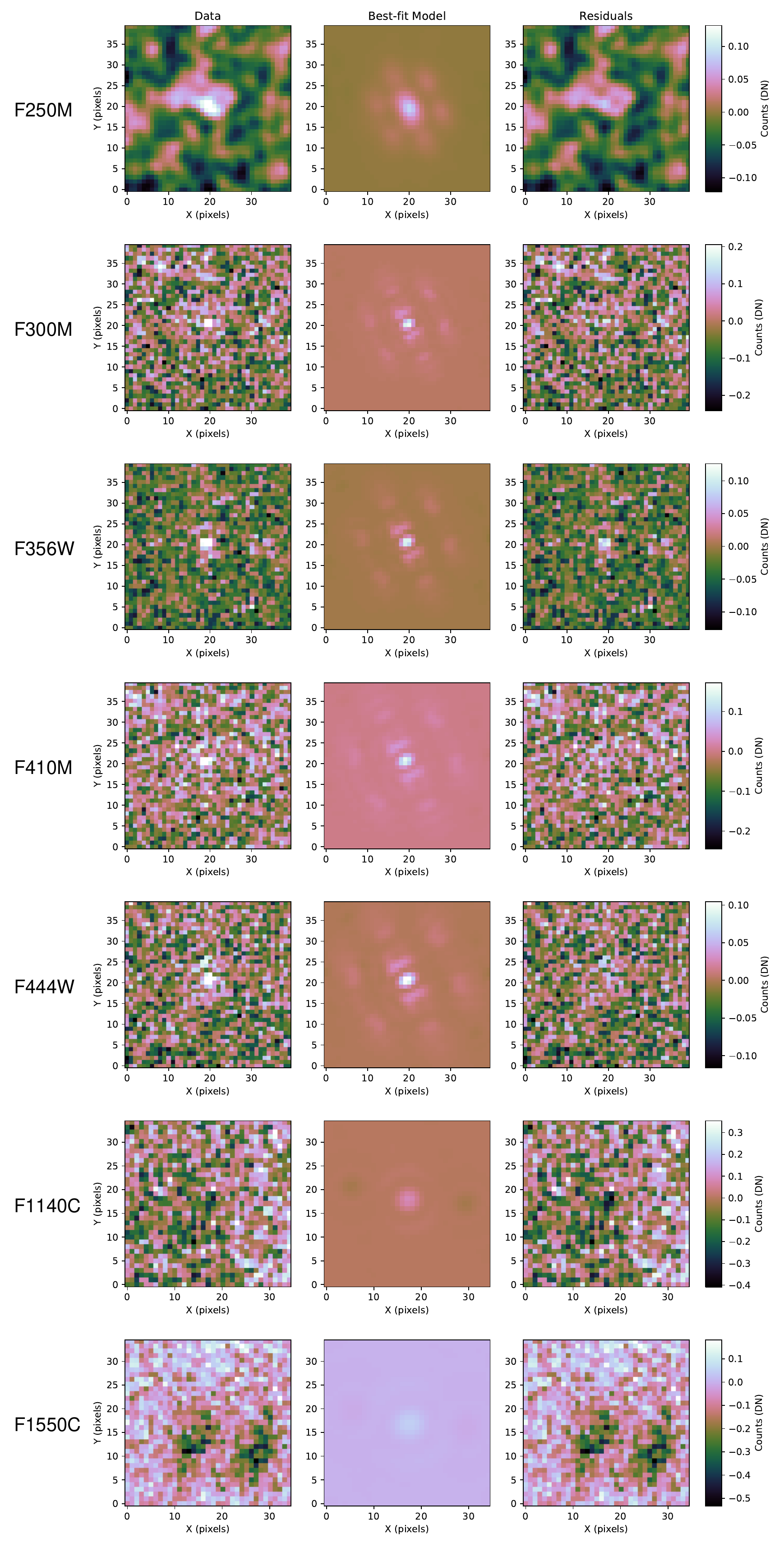}{0.5\linewidth}{(c11)}
    \fig{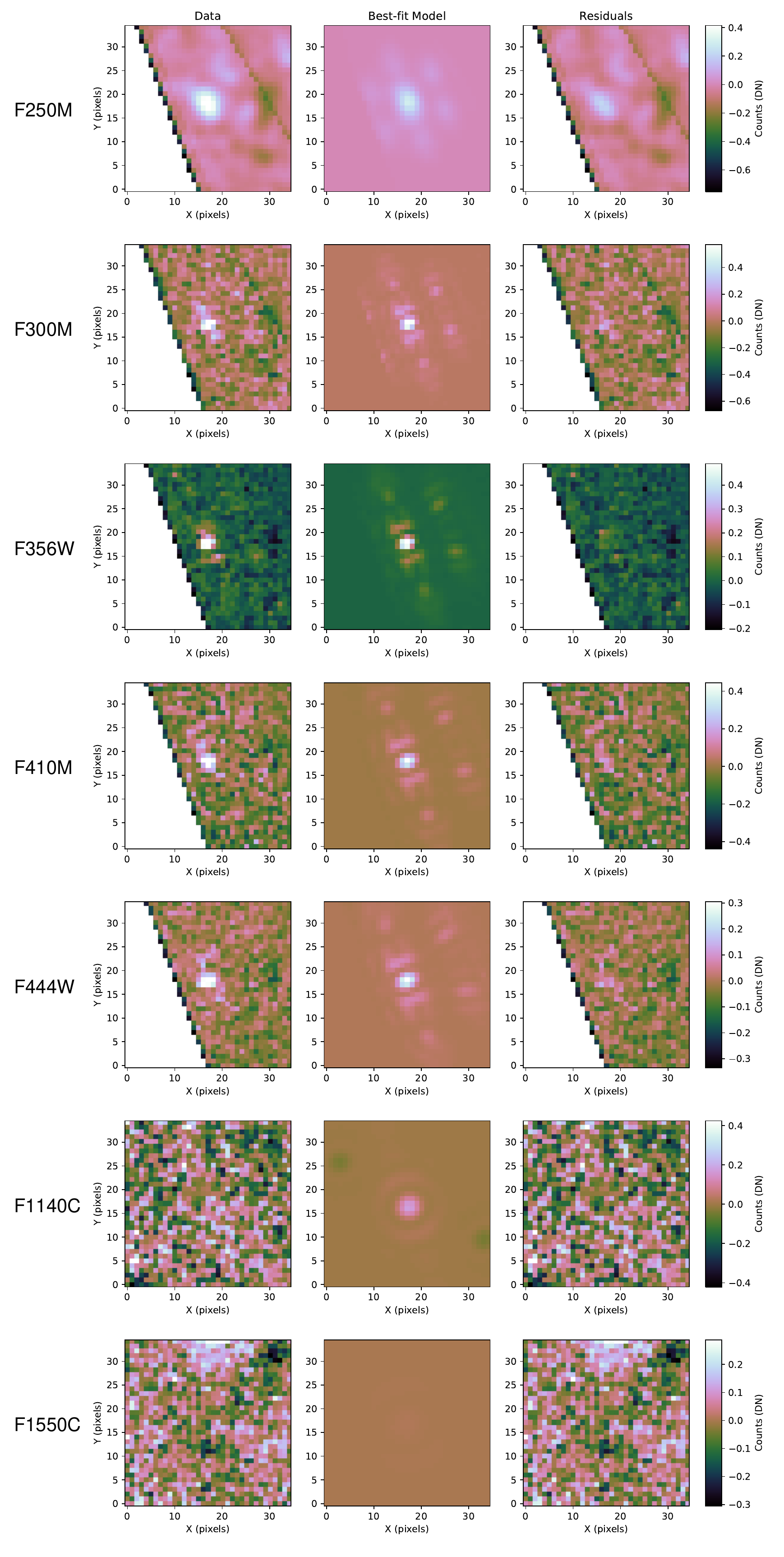}{0.5\linewidth}{(c12)}
    }
    \caption{Companion fits for c11 (using the Powell minimization method) and c12.}
    \label{fig:psf_c11c12}
\end{figure}
\begin{figure}
    \figurenum{C.7}
    \centering
    \gridline{
    \fig{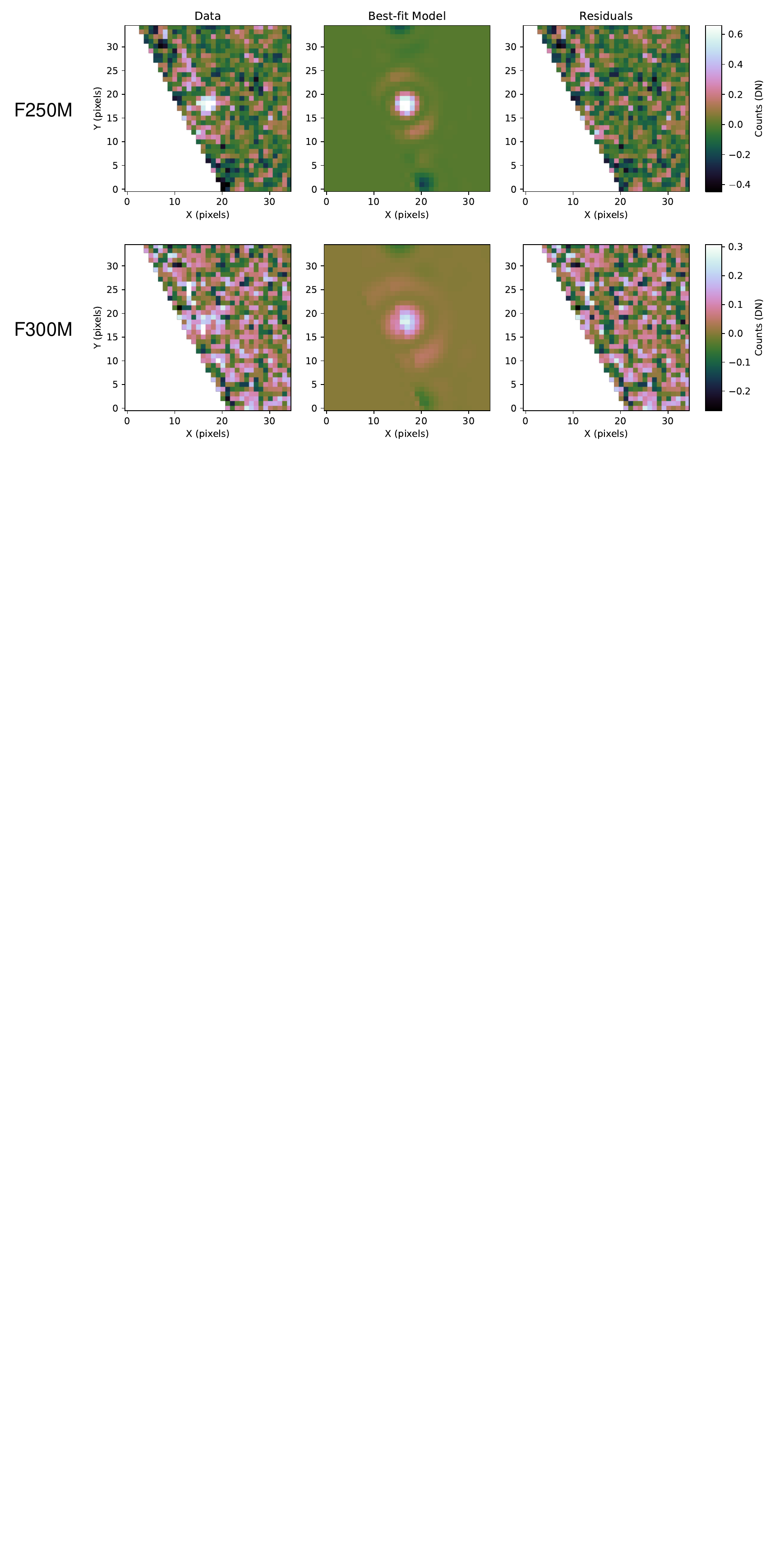}{0.5\linewidth}{(c18)}
    \fig{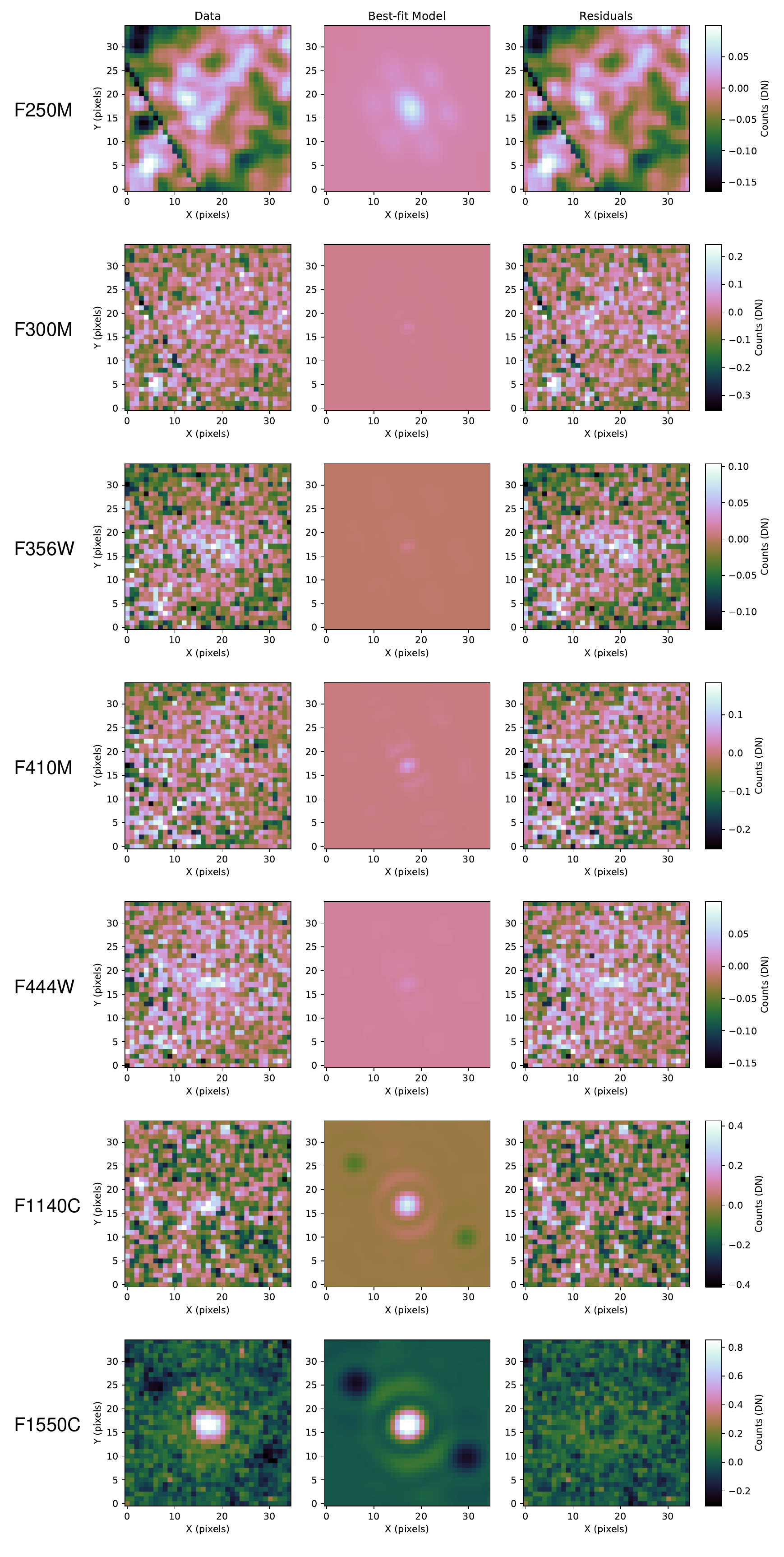}{0.5\linewidth}{(c20)}}
    \caption{Companion fits for c18 and c20 using the Powell minimization method. c18 was outside of the field of view in NIRCam and therefore is only observed in F1140C and F1550C.}
    \label{fig:psf_c18c20}
\end{figure} 

\clearpage

\startlongtable
\begin{deluxetable*}{lllllll}

\tablecaption{JWST astrometry and photometry of all vetted background sources. The position angle ($\theta$) is provided from North through East. All apparent magnitudes are relative to Vega. Not all sources were detected in all filters, so in some cases, the error may be larger than the measured flux value and should thus be treated as a $1\sigma$ upper limit. Sources are marked with an s, g, or ? depending on whether it was classified as a star, galaxy, or unknown (in the case of c18).\label{tab:sources_table}}

\tablehead{
\colhead{Source} &
\colhead{Filter} &
\colhead{$\rho$ (mas)} &
\colhead{$\theta$ (deg)} &
\colhead{Flux (Jy)} &
\colhead{Flux (W m$^{-2}$ $\mu$m$^{-1}$)} &
\colhead{$m_b$ (mag)}
}

\startdata \\
c1 (s) & F250M & 9324 $\pm$ 7 & 12.1 $\pm$ 0.04 & (1.64 $\pm$ 0.18) $\times$ 10$^{-5}$ & (7.86 $\pm$ 0.87) $\times$ 10$^{-18}$ & 18.719 $\pm$ 0.113 \\
 & F300M & 9320 $\pm$ 6 & 12.2 $\pm$ 0.04 & (1.25 $\pm$ 0.22) $\times$ 10$^{-5}$ & (4.17 $\pm$ 0.73) $\times$ 10$^{-18}$ & 18.671 $\pm$ 0.168 \\
 & F356W & 9315 $\pm$ 6 & 12.2 $\pm$ 0.04 & (1.02 $\pm$ 0.09) $\times$ 10$^{-5}$ & (2.40 $\pm$ 0.20) $\times$ 10$^{-18}$ & 18.558 $\pm$ 0.095 \\
 & F410M & 9320 $\pm$ 7 & 12.1 $\pm$ 0.04 & (9.04 $\pm$ 2.13) $\times$ 10$^{-6}$ & (1.63 $\pm$ 0.39) $\times$ 10$^{-18}$ & 18.413 $\pm$ 0.239 \\
 & F444W & 9318 $\pm$ 6 & 12.1 $\pm$ 0.04 & (8.03 $\pm$ 2.11) $\times$ 10$^{-6}$ & (1.26 $\pm$ 0.33) $\times$ 10$^{-18}$ & 18.418 $\pm$ 0.330 \\
 & F1140C & 9413 $\pm$ 28 & 12.0 $\pm$ 0.29 &  $<$(6.04 $\pm$ 1.58) $\times$ 10$^{-6}$ & $<$(1.42 $\pm$ 0.37) $\times$ 10$^{-19}$ & $>$16.727 $\pm$ 0.284 \\
 & F1550C & 9328 $\pm$ 77 & 12.3 $\pm$ 0.45 &  $<$(4.18 $\pm$ 4.36) $\times$ 10$^{-7}$ & $<$(5.21 $\pm$ 5.43) $\times$ 10$^{-21}$ & $>$18.954 $\pm$ 1.132 \\
\midrule
c2 (s) & F250M & 8832 $\pm$ 6 & 343.7 $\pm$ 0.04 & (4.53 $\pm$ 0.32) $\times$ 10$^{-5}$ & (2.17 $\pm$ 0.15) $\times$ 10$^{-17}$ & 17.616 $\pm$ 0.078 \\
 & F300M & 8840 $\pm$ 6 & 343.7 $\pm$ 0.04 & (3.26 $\pm$ 0.22) $\times$ 10$^{-5}$ & (1.09 $\pm$ 0.07) $\times$ 10$^{-17}$ & 17.628 $\pm$ 0.073 \\
 & F356W & 8892 $\pm$ 6 & 343.8 $\pm$ 0.04 & (1.85 $\pm$ 0.16) $\times$ 10$^{-5}$ & (4.36 $\pm$ 0.38) $\times$ 10$^{-18}$ & 17.912 $\pm$ 0.073 \\
 & F410M & 8842 $\pm$ 6 & 343.7 $\pm$ 0.04 & (2.00 $\pm$ 0.14) $\times$ 10$^{-5}$ & (3.61 $\pm$ 0.26) $\times$ 10$^{-18}$ & 17.552 $\pm$ 0.077 \\
 & F444W & 8882 $\pm$ 6 & 343.7 $\pm$ 0.04 & (1.48 $\pm$ 0.12) $\times$ 10$^{-5}$ & (2.33 $\pm$ 0.19) $\times$ 10$^{-18}$ & 17.753 $\pm$ 0.078 \\
 & F1140C & 8845 $\pm$ 15 & 342.9 $\pm$ 0.18 & (3.17 $\pm$ 0.64) $\times$ 10$^{-6}$ & (7.44 $\pm$ 1.49) $\times$ 10$^{-20}$ & 17.428 $\pm$ 0.218 \\
 & F1550C & 8932 $\pm$ 88 & 343.6 $\pm$ 0.51 & (8.18 $\pm$ 4.85) $\times$ 10$^{-7}$ & (1.02 $\pm$ 0.60) $\times$ 10$^{-20}$ & 18.226 $\pm$ 0.644 \\
\midrule
c3 (s) & F250M & 8126 $\pm$ 6 & 337.0 $\pm$ 0.04 & (9.05 $\pm$ 0.64) $\times$ 10$^{-5}$ & (4.33 $\pm$ 0.31) $\times$ 10$^{-17}$ & 16.866 $\pm$ 0.077 \\
 & F300M & 8144 $\pm$ 6 & 337.1 $\pm$ 0.04 & (5.94 $\pm$ 0.41) $\times$ 10$^{-5}$ & (1.98 $\pm$ 0.14) $\times$ 10$^{-17}$ & 16.978 $\pm$ 0.075 \\
 & F356W & 8148 $\pm$ 6 & 337.2 $\pm$ 0.04 & (4.94 $\pm$ 0.33) $\times$ 10$^{-5}$ & (1.16 $\pm$ 0.08) $\times$ 10$^{-17}$ & 16.848 $\pm$ 0.073 \\
 & F410M & 8144 $\pm$ 6 & 337.1 $\pm$ 0.04 & (4.11 $\pm$ 0.28) $\times$ 10$^{-5}$ & (7.44 $\pm$ 0.51) $\times$ 10$^{-18}$ & 16.768 $\pm$ 0.075 \\
 & F444W & 8143 $\pm$ 6 & 337.1 $\pm$ 0.04 & (3.56 $\pm$ 0.25) $\times$ 10$^{-5}$ & (5.60 $\pm$ 0.40) $\times$ 10$^{-18}$ & 16.802 $\pm$ 0.077 \\
 & F1140C & 8168 $\pm$ 20 & 336.9 $\pm$ 0.14 & (5.27 $\pm$ 0.91) $\times$ 10$^{-6}$ & (1.24 $\pm$ 0.21) $\times$ 10$^{-19}$ & 16.877 $\pm$ 0.188 \\
 & F1550C & 8215 $\pm$ 27 & 337.1 $\pm$ 0.20 & (3.62 $\pm$ 0.75) $\times$ 10$^{-6}$ & (4.51 $\pm$ 0.93) $\times$ 10$^{-20}$ & 16.610 $\pm$ 0.224 \\
\midrule
c4 (s) & F250M & 8605 $\pm$ 6 & 322.5 $\pm$ 0.04 & (8.86 $\pm$ 0.63) $\times$ 10$^{-5}$ & (4.24 $\pm$ 0.30) $\times$ 10$^{-17}$ & 16.888 $\pm$ 0.077 \\
 & F300M & 8607 $\pm$ 6 & 322.7 $\pm$ 0.04 & (6.60 $\pm$ 0.45) $\times$ 10$^{-5}$ & (2.20 $\pm$ 0.15) $\times$ 10$^{-17}$ & 16.864 $\pm$ 0.074 \\
 & F356W & 8607 $\pm$ 6 & 322.6 $\pm$ 0.04 & (5.00 $\pm$ 0.34) $\times$ 10$^{-5}$ & (1.18 $\pm$ 0.08) $\times$ 10$^{-17}$ & 16.834 $\pm$ 0.073 \\
 & F410M & 8611 $\pm$ 6 & 322.6 $\pm$ 0.04 & (4.21 $\pm$ 0.29) $\times$ 10$^{-5}$ & (7.60 $\pm$ 0.52) $\times$ 10$^{-18}$ & 16.744 $\pm$ 0.075 \\
 & F444W & 8611 $\pm$ 6 & 322.6 $\pm$ 0.04 & (3.53 $\pm$ 0.25) $\times$ 10$^{-5}$ & (5.56 $\pm$ 0.39) $\times$ 10$^{-18}$ & 16.809 $\pm$ 0.077 \\
 & F1140C & 8699 $\pm$ 20 & 322.6 $\pm$ 0.13 & (5.25 $\pm$ 0.91) $\times$ 10$^{-6}$ & (1.23 $\pm$ 0.21) $\times$ 10$^{-19}$ & 16.880 $\pm$ 0.189 \\
 & F1550C & 8646 $\pm$ 44 & 322.7 $\pm$ 0.31 & (2.43 $\pm$ 0.60) $\times$ 10$^{-6}$ & (3.03 $\pm$ 0.74) $\times$ 10$^{-20}$ & 17.043 $\pm$ 0.266 \\
\midrule
c5 (s) & F250M & 6631 $\pm$ 6 & 306.0 $\pm$ 0.06 & (1.22 $\pm$ 0.09) $\times$ 10$^{-5}$ & (5.84 $\pm$ 0.43) $\times$ 10$^{-18}$ & 19.041 $\pm$ 0.081 \\
 & F300M & 6634 $\pm$ 6 & 306.2 $\pm$ 0.06 & (9.30 $\pm$ 0.71) $\times$ 10$^{-6}$ & (3.10 $\pm$ 0.24) $\times$ 10$^{-18}$ & 18.991 $\pm$ 0.084 \\
 & F356W & 6631 $\pm$ 6 & 306.2 $\pm$ 0.05 & (7.64 $\pm$ 0.51) $\times$ 10$^{-6}$ & (1.80 $\pm$ 0.12) $\times$ 10$^{-18}$ & 18.874 $\pm$ 0.073 \\
 & F410M & 6631 $\pm$ 7 & 306.2 $\pm$ 0.06 & (5.86 $\pm$ 0.45) $\times$ 10$^{-6}$ & (1.06 $\pm$ 0.08) $\times$ 10$^{-18}$ & 18.884 $\pm$ 0.083 \\
 & F444W & 6634 $\pm$ 6 & 306.2 $\pm$ 0.06 & (5.03 $\pm$ 0.39) $\times$ 10$^{-6}$ & (7.93 $\pm$ 0.61) $\times$ 10$^{-19}$ & 18.925 $\pm$ 0.084 \\
 & F1140C & 6660 $\pm$ 51 & 306.3 $\pm$ 0.44 &  $<$(9.55 $\pm$ 4.99) $\times$ 10$^{-7}$ & $<$(2.24 $\pm$ 1.17) $\times$ 10$^{-20}$ & $>$18.731 $\pm$ 0.567 \\
 & F1550C & 6630 $\pm$ 75 & 306.1 $\pm$ 0.65 &  $<$(1.26 $\pm$ 1.30) $\times$ 10$^{-7}$ & $<$(1.57 $\pm$ 1.62) $\times$ 10$^{-21}$ & $>$20.256 $\pm$ 1.123 \\
\midrule
c6 (s) & F250M & 10632 $\pm$ 6 & 288.0 $\pm$ 0.03 & (5.05 $\pm$ 1.84) $\times$ 10$^{-5}$ & (2.41 $\pm$ 0.88) $\times$ 10$^{-17}$ & 17.500 $\pm$ 0.286 \\
 & F300M & 10594 $\pm$ 6 & 288.1 $\pm$ 0.03 & (3.61 $\pm$ 1.26) $\times$ 10$^{-5}$ & (1.20 $\pm$ 0.42) $\times$ 10$^{-17}$ & 17.518 $\pm$ 0.289 \\
 & F356W & 10597 $\pm$ 6 & 288.1 $\pm$ 0.03 & (3.41 $\pm$ 1.80) $\times$ 10$^{-5}$ & (8.03 $\pm$ 4.23) $\times$ 10$^{-18}$ & 17.249 $\pm$ 0.864 \\
 & F410M & 10600 $\pm$ 6 & 288.1 $\pm$ 0.03 & (3.03 $\pm$ 1.62) $\times$ 10$^{-5}$ & (5.48 $\pm$ 2.93) $\times$ 10$^{-18}$ & 17.100 $\pm$ 1.262 \\
 & F444W & 10605 $\pm$ 6 & 288.1 $\pm$ 0.03 & (2.60 $\pm$ 1.27) $\times$ 10$^{-5}$ & (4.10 $\pm$ 2.00) $\times$ 10$^{-18}$ & 17.141 $\pm$ 0.634 \\
 & F1140C & 10579 $\pm$ 37 & 288.5 $\pm$ 0.23 & (1.26 $\pm$ 0.54) $\times$ 10$^{-6}$ & (2.97 $\pm$ 1.28) $\times$ 10$^{-20}$ & 18.426 $\pm$ 0.468 \\
 & F1550C & 10659 $\pm$ 67 & 288.3 $\pm$ 0.35 &  $<$(5.79 $\pm$ 4.60) $\times$ 10$^{-7}$ & $<$(7.22 $\pm$ 5.73) $\times$ 10$^{-21}$ & $>$18.600 $\pm$ 0.862 \\
\midrule
c8 (s) & F250M & 7554 $\pm$ 8 & 252.5 $\pm$ 0.06 & (4.21 $\pm$ 0.48) $\times$ 10$^{-6}$ & (2.01 $\pm$ 0.23) $\times$ 10$^{-18}$ & 20.197 $\pm$ 0.116 \\
 & F300M & 7550 $\pm$ 9 & 252.5 $\pm$ 0.06 & (2.81 $\pm$ 0.35) $\times$ 10$^{-6}$ & (9.38 $\pm$ 1.17) $\times$ 10$^{-19}$ & 20.290 $\pm$ 0.120 \\
 & F356W & 7550 $\pm$ 8 & 252.6 $\pm$ 0.06 & (2.92 $\pm$ 0.23) $\times$ 10$^{-6}$ & (6.87 $\pm$ 0.53) $\times$ 10$^{-19}$ & 19.918 $\pm$ 0.085 \\
 & F410M & 7556 $\pm$ 10 & 252.5 $\pm$ 0.07 & (2.90 $\pm$ 0.33) $\times$ 10$^{-6}$ & (5.24 $\pm$ 0.60) $\times$ 10$^{-19}$ & 19.649 $\pm$ 0.119 \\
 & F444W & 7542 $\pm$ 10 & 252.6 $\pm$ 0.07 & (2.28 $\pm$ 0.26) $\times$ 10$^{-6}$ & (3.60 $\pm$ 0.42) $\times$ 10$^{-19}$ & 19.783 $\pm$ 0.132 \\
 & F1140C & 7559 $\pm$ 75 & 252.7 $\pm$ 0.58 & $<$(1.07 $\pm$ 1.11) $\times$ 10$^{-7}$ & $<$(2.51 $\pm$ 2.60) $\times$ 10$^{-21}$ & $>$21.109 $\pm$ 1.124 \\
 & F1550C & 7570 $\pm$ 74 & 252.7 $\pm$ 0.58 & $<$(1.17 $\pm$ 0.97) $\times$ 10$^{-7}$ & $<$(1.46 $\pm$ 1.21) $\times$ 10$^{-21}$ & $>$20.337 $\pm$ 0.900 \\
\midrule
c9 (g) & F250M & 7868 $\pm$ 42 & 242.4 $\pm$ 0.35 & (9.02 $\pm$ 7.01) $\times$ 10$^{-7}$ & (4.31 $\pm$ 3.35) $\times$ 10$^{-19}$ & 21.870 $\pm$ 1.649 \\
 & F300M & 7962 $\pm$ 69 & 242.0 $\pm$ 0.48 & (4.96 $\pm$ 0.78) $\times$ 10$^{-7}$ & (1.65 $\pm$ 0.26) $\times$ 10$^{-19}$ & 22.174 $\pm$ 0.649 \\
 & F356W & 7947 $\pm$ 49 & 242.0 $\pm$ 0.29 & (4.58 $\pm$ 2.85) $\times$ 10$^{-7}$ & (1.08 $\pm$ 0.67) $\times$ 10$^{-19}$ & 21.930 $\pm$ 1.157 \\
 & F410M & 7977 $\pm$ 61 & 242.0 $\pm$ 0.43 & (4.93 $\pm$ 1.94) $\times$ 10$^{-7}$ & (8.92 $\pm$ 3.51) $\times$ 10$^{-20}$ & 21.571 $\pm$ 1.027 \\
 & F444W & 7950 $\pm$ 48 & 242.2 $\pm$ 0.34 & (4.63 $\pm$ 0.49) $\times$ 10$^{-7}$ & (7.29 $\pm$ 0.77) $\times$ 10$^{-20}$ & 21.517 $\pm$ 0.557 \\
 & F1140C & 7873 $\pm$ 66 & 242.4 $\pm$ 0.41 & $<$(9.31 $\pm$ 3.78) $\times$ 10$^{-7}$ & $<$(2.18 $\pm$ 0.89) $\times$ 10$^{-20}$ & $>$18.759 $\pm$ 0.441 \\
 & F1550C & 7964 $\pm$ 75 & 241.9 $\pm$ 0.52 & $<$(1.16 $\pm$ 0.38) $\times$ 10$^{-6}$ & $<$(1.45 $\pm$ 0.47) $\times$ 10$^{-20}$ & $>$17.845 $\pm$ 0.356 \\
\midrule
c10 (g) & F250M & 6729 $\pm$ 19 & 154.6 $\pm$ 0.22 & (8.63 $\pm$ 3.85) $\times$ 10$^{-7}$ & (4.13 $\pm$ 1.84) $\times$ 10$^{-19}$ & 21.918 $\pm$ 0.351 \\
 & F300M & 6699 $\pm$ 18 & 154.5 $\pm$ 0.18 & (7.28 $\pm$ 3.90) $\times$ 10$^{-7}$ & (2.43 $\pm$ 1.30) $\times$ 10$^{-19}$ & 21.758 $\pm$ 1.058 \\
 & F356W & 6718 $\pm$ 16 & 154.6 $\pm$ 0.16 & (5.85 $\pm$ 2.16) $\times$ 10$^{-7}$ & (1.37 $\pm$ 0.51) $\times$ 10$^{-19}$ & 21.664 $\pm$ 0.248 \\
 & F410M & 6704 $\pm$ 16 & 154.5 $\pm$ 0.16 & (6.61 $\pm$ 2.79) $\times$ 10$^{-7}$ & (1.19 $\pm$ 0.50) $\times$ 10$^{-19}$ & 21.254 $\pm$ 0.306 \\
 & F444W & 6695 $\pm$ 13 & 154.7 $\pm$ 0.13 & (7.62 $\pm$ 1.56) $\times$ 10$^{-7}$ & (1.20 $\pm$ 0.25) $\times$ 10$^{-19}$ & 20.975 $\pm$ 0.226 \\
 & F1140C & 6683 $\pm$ 64 & 153.9 $\pm$ 0.60 & $<$(9.83 $\pm$ 3.75) $\times$ 10$^{-7}$ & $<$(2.31 $\pm$ 0.88) $\times$ 10$^{-20}$ & $>$18.699 $\pm$ 0.414 \\
 & F1550C & 6700 $\pm$ 72 & 154.1 $\pm$ 0.63 & $<$(1.13 $\pm$ 0.29) $\times$ 10$^{-6}$ & $<$(1.41 $\pm$ 0.36) $\times$ 10$^{-20}$ & $>$17.875 $\pm$ 0.276 \\
\midrule
c11 (g) & F250M & 8202 $\pm$ 67 & 172.0 $\pm$ 0.41 & (6.14 $\pm$ 3.88) $\times$ 10$^{-7}$ & (2.94 $\pm$ 1.86) $\times$ 10$^{-19}$ & 22.287 $\pm$ 1.202 \\
 & F300M & 8152 $\pm$ 67 & 171.7 $\pm$ 0.48 & (5.32 $\pm$ 1.25) $\times$ 10$^{-7}$ & (1.77 $\pm$ 0.42) $\times$ 10$^{-19}$ & 22.098 $\pm$ 0.687 \\
 & F356W & 8135 $\pm$ 67 & 171.6 $\pm$ 0.57 & (3.99 $\pm$ 0.39) $\times$ 10$^{-7}$ & (9.38 $\pm$ 0.92) $\times$ 10$^{-20}$ & 22.079 $\pm$ 0.514 \\
 & F410M & 8136 $\pm$ 47 & 171.4 $\pm$ 0.50 & (6.21 $\pm$ 1.19) $\times$ 10$^{-7}$ & (1.12 $\pm$ 0.22) $\times$ 10$^{-19}$ & 21.322 $\pm$ 0.894 \\
 & F444W & 8134 $\pm$ 58 & 171.6 $\pm$ 0.50 & (5.29 $\pm$ 0.88) $\times$ 10$^{-7}$ & (8.33 $\pm$ 1.38) $\times$ 10$^{-20}$ & 21.371 $\pm$ 0.828 \\
 & F1140C & 8116 $\pm$ 79 & 171.6 $\pm$ 0.57 & $<$(7.42 $\pm$ 1.83) $\times$ 10$^{-7}$ & $<$(1.74 $\pm$ 0.43) $\times$ 10$^{-20}$ & $>$19.005 $\pm$ 0.268 \\
 & F1550C & 8131 $\pm$ 74 & 171.7 $\pm$ 0.55 & $<$(7.55 $\pm$ 1.84) $\times$ 10$^{-7}$ & $<$(9.40 $\pm$ 2.29) $\times$ 10$^{-21}$ & $>$18.313 $\pm$ 0.264 \\
\midrule
c12 (s) & F250M & 11552 $\pm$ 10 & 151.9 $\pm$ 0.05 & (4.11 $\pm$ 2.67) $\times$ 10$^{-6}$ & (1.97 $\pm$ 1.28) $\times$ 10$^{-18}$ & 20.222 $\pm$ 1.266 \\
 & F300M & 11584 $\pm$ 9 & 152.0 $\pm$ 0.05 & (3.70 $\pm$ 1.48) $\times$ 10$^{-6}$ & (1.23 $\pm$ 0.49) $\times$ 10$^{-18}$ & 19.993 $\pm$ 1.103 \\
 & F356W & 11598 $\pm$ 7 & 151.8 $\pm$ 0.04 & (3.98 $\pm$ 2.46) $\times$ 10$^{-6}$ & (9.36 $\pm$ 5.79) $\times$ 10$^{-19}$ & 19.581 $\pm$ 1.273 \\
 & F410M & 11587 $\pm$ 9 & 151.9 $\pm$ 0.05 & (3.40 $\pm$ 1.48) $\times$ 10$^{-6}$ & (6.14 $\pm$ 2.67) $\times$ 10$^{-19}$ & 19.476 $\pm$ 0.929 \\
 & F444W & 11588 $\pm$ 8 & 151.9 $\pm$ 0.05 & (2.75 $\pm$ 1.11) $\times$ 10$^{-6}$ & (4.33 $\pm$ 1.75) $\times$ 10$^{-19}$ & 19.582 $\pm$ 0.976 \\
 & F1140C & 11608 $\pm$ 55 & 152.4 $\pm$ 0.19 & $<$(1.56 $\pm$ 0.54) $\times$ 10$^{-6}$ & $<$(3.66 $\pm$ 1.27) $\times$ 10$^{-20}$ & $>$18.197 $\pm$ 0.377 \\
 & F1550C & 11616 $\pm$ 74 & 152.1 $\pm$ 0.38 & $<$(1.01 $\pm$ 0.81) $\times$ 10$^{-7}$ & $<$(1.26 $\pm$ 1.01) $\times$ 10$^{-21}$ & $>$20.498 $\pm$ 0.875 \\
\midrule
c18 (?) & F1140C & 11534 $\pm$ 16 & 97.5 $\pm$ 0.08 & (7.34 $\pm$ 1.25) $\times$ 10$^{-6}$ & (1.72 $\pm$ 0.29) $\times$ 10$^{-19}$ & 16.516 $\pm$ 0.185 \\
 & F1550C & 11571 $\pm$ 35 & 97.4 $\pm$ 0.15 & (4.70 $\pm$ 0.97) $\times$ 10$^{-6}$ & (5.85 $\pm$ 1.21) $\times$ 10$^{-20}$ & 16.328 $\pm$ 0.225 \\
\midrule
c20 (g) & F250M & 9473 $\pm$ 36 & 146.6 $\pm$ 0.25 &  $<$(5.18 $\pm$ 3.58) $\times$ 10$^{-7}$ &  $<$(2.48 $\pm$ 1.71) $\times$ 10$^{-19}$ & $>$22.471 $\pm$ 1.725 \\
 & F300M & 9488 $\pm$ 42 & 146.3 $\pm$ 0.26 &  $<$(5.08 $\pm$ 20.96) $\times$ 10$^{-8}$ &  $<$(1.69 $\pm$ 6.99) $\times$ 10$^{-20}$ & $>$24.648 $\pm$ 1.150 \\
 & F356W & 9492 $\pm$ 44 & 146.3 $\pm$ 0.28 &  $<$(3.58 $\pm$ 27.28) $\times$ 10$^{-8}$ &  $<$(8.43 $\pm$ 64.15) $\times$ 10$^{-21}$ & $>$24.696 $\pm$ 1.362 \\
 & F410M & 9499 $\pm$ 35 & 146.3 $\pm$ 0.22 &  $<$(2.35 $\pm$ 2.77) $\times$ 10$^{-7}$ &  $<$(4.25 $\pm$ 5.00) $\times$ 10$^{-20}$ & $>$22.377 $\pm$ 1.089 \\
 & F444W & 9490 $\pm$ 41 & 146.2 $\pm$ 0.25 &  $<$(4.98 $\pm$ 19.19) $\times$ 10$^{-8}$ &  $<$(7.84 $\pm$ 30.22) $\times$ 10$^{-21}$ & $>$23.937 $\pm$ 1.269 \\
 & F1140C & 9538 $\pm$ 45 & 146.2 $\pm$ 0.27 &  (3.07 $\pm$ 0.93) $\times$ 10$^{-6}$ &  (7.19 $\pm$ 2.18) $\times$ 10$^{-20}$ & 17.464 $\pm$ 0.329 \\
 & F1550C & 9555 $\pm$ 17 & 146.4 $\pm$ 0.11 &  (1.61 $\pm$ 0.29) $\times$ 10$^{-5}$ & (2.01 $\pm$ 0.37) $\times$ 10$^{-19}$ &  14.991 $\pm$ 0.198 \\
\enddata

\end{deluxetable*}

\setcounter{figure}{0} 
\setcounter{table}{0} 

\section{Companion Candidates} \label{sec:companion_candidates}
\renewcommand{\thefigure}{D.\arabic{figure}}
\renewcommand{\theHfigure}{D.\arabic{figure}}
\renewcommand{\thetable}{D.\arabic{table}}
\renewcommand{\theHtable}{D.\arabic{table}}
\begin{figure*}[h!]
    \centering
    \includegraphics[width=0.88\linewidth]{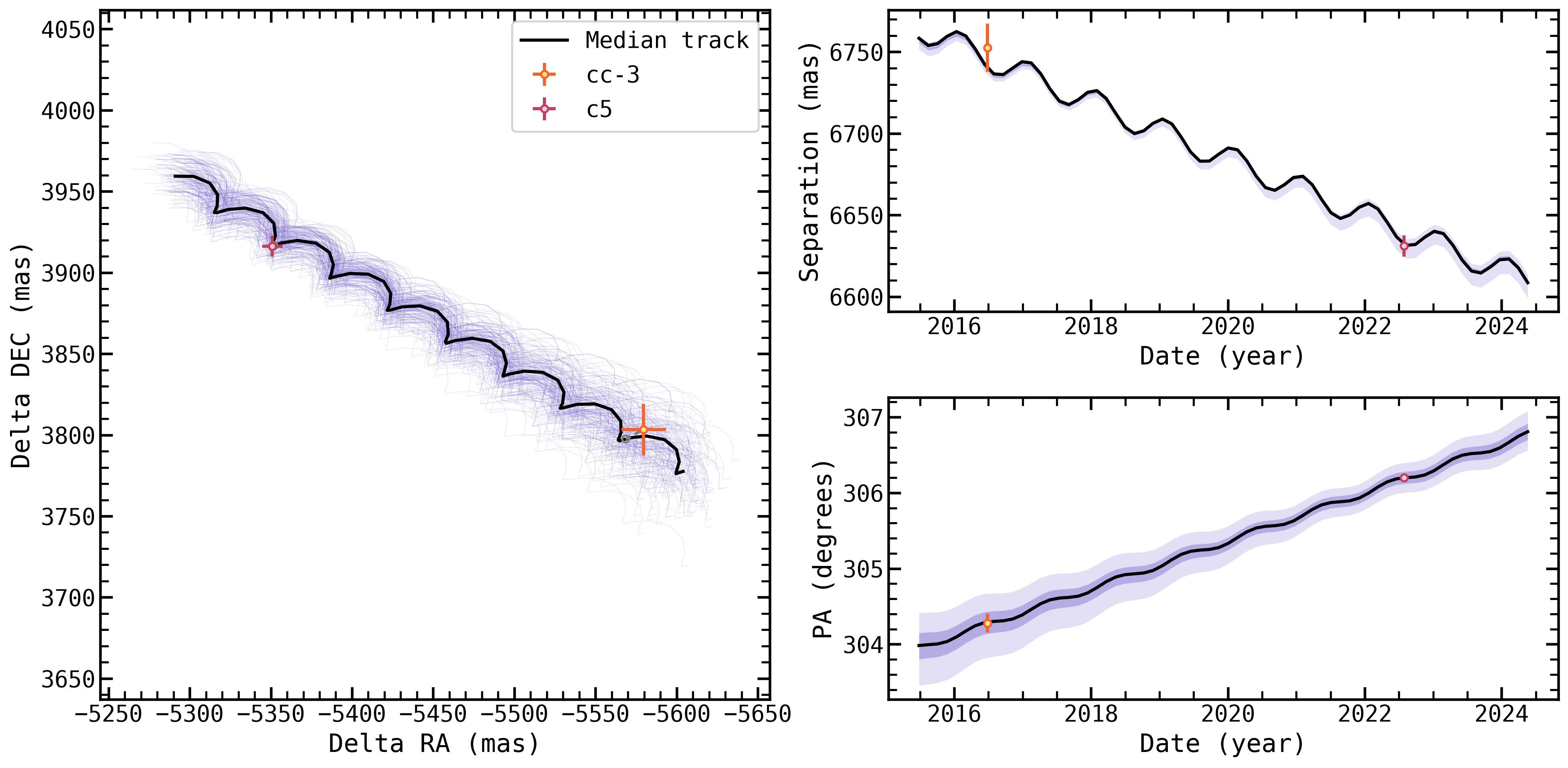}
    \caption{The backtracks analysis that correlates c5 with cc-3 from \cite{chauvin2017}.}
    \label{fig:backtracks_c5}
\end{figure*}
\end{document}